\documentclass[notitlepage,aps,pra,twoside,superscriptaddress,showkeys,nofootinbib]{revtex4-1}
 \usepackage{graphicx}
 \usepackage{amsmath,amsfonts,amssymb,amsthm,amsbsy,mathtools}
 \usepackage{blindtext}
 \usepackage[utf8]{inputenc}
 \usepackage[T1]{fontenc}
 \usepackage{amssymb} \usepackage{color,graphicx} \usepackage{amsmath}
 \usepackage{amsbsy} \usepackage{amsthm} 
 \usepackage{bm} \usepackage{float} 
 \usepackage{placeins,cals}
 \usepackage{setspace,ulem}
 \usepackage[colorlinks=true,allcolors=blue!80!black]{hyperref}
 \usepackage{xfrac}
 \usepackage{amsfonts}
 \usepackage{latexsym}
 \usepackage{amsmath,amssymb}
 \usepackage{mathrsfs}
 \usepackage{color,xcolor}
 \usepackage{slashed}
 \usepackage{url}
 \usepackage{empheq}
 \usepackage{textcomp}
 \usepackage{dcolumn}
 \usepackage{ulem}
 \usepackage{subfigure}
 \usepackage{amsfonts}
 \usepackage{epsfig}
 \usepackage{bbm}
 \usepackage{tabularx}
 \usepackage{multirow}
 \usepackage{mathtools}
 \usepackage{xfrac}
 \usepackage{epsfig}
 \usepackage{amsfonts}
 \usepackage{latexsym}
 \usepackage{mathrsfs}
 \usepackage{hyperref}
 \usepackage{setspace}
 \usepackage{color}
 \usepackage{bm}
 \usepackage{orcidlink}
 \usepackage{isotope}
 \usepackage{braket}
 \usepackage{braket}
 \newcommand{\bq}{\begin{equation}} \newcommand{\eq}{\end{equation}}
 \newcommand{\bqali}{\begin{equation}\begin{aligned}} \newcommand{\eqali}{\end{aligned}\end{equation}}

 \def \x{\mathbf{ x}}

 \def\comment#1{}

 \def \k{\mathbf{ k}}
 \def \p{\mathbf{ p}}
 \def \q{\mathbf{ q}}

 \def \l{\left}
 \def \r{\right}
 \def\beq{\begin{equation}}
 	\def\eeq{\end{equation}}
 \def\bea{\begin{eqnarray}}
 	\def\eea{\end{eqnarray}}
 \usepackage{mhchem}
 
 \def\inte{\textrm{int}}

 \def\k{\textbf{k}}

 \allowdisplaybreaks
 \makeatletter
 \usepackage{cancel}
 \usepackage{color}

 \makeatletter
 \providecommand\NAT@force@numbers{}

 \makeatother

\begin{document}

\title{	Microscopic Origins of Collapse Models: Decoherence from Graviton Bremsstrahlung }

\author{M. Zarei\,\orcidlink{0000-0001-7744-2817}
}
\email[]{m.zarei@iut.ac.ir}
\affiliation{Department of Physics, Isfahan University of Technology, 84156-83111 Isfahan, Iran}
\affiliation{Quantum Technology Research Group, Isfahan University of Technology, 84156-83111 Isfahan, Iran}

\date{\today}

	 \begin{abstract}
	 	Some collapse models propose that gravitational effects cause the instability of
	 	spatial superpositions of mass distributions, leading to wave function collapse.
	 	In this paper, we utilize the quantum Boltzmann equation~(QBE) to analyze the
	 	decoherence of a spin-$1/2$ fermion in a spatial superposition under graviton
	 	bremsstrahlung emission. We introduce a quantitative measure linking the stability
	 	of the superposition to the spatial separation $\Delta\mathbf{x}$, the fermion
	 	mass $m_f$, and the gravitational coupling $\kappa$. By examining the collision
	 	term in the QBE, we derive the decoherence rate $\Gamma(\Delta\mathbf{x})$ and
	 	show how it depends on these parameters. Our results provide a detailed framework
	 	for understanding graviton induced decoherence, bridging the gap between quantum
	 	field theory and collapse models. We also discuss the implications of these
	 	findings for experimental tests of gravitationally induced wave function collapse
	 	and the broader class of collapse models known as the dissipative continuous
	 	spontaneous localization~(CSL) model.
	 \end{abstract}

\maketitle

\maketitle

\section{Introduction} 

   Collapse models, such as the dissipative CSL model \cite{Smirne:2014paa}, represent some of the most developed theoretical frameworks addressing the quantum measurement problem \cite{Tilloy:2015zya,Kafri:2014zsa}. Moreover, the Diósi-Penrose (DP) model \cite{Diosi:1989hlx,Penrose:1996cv} proposes a gravity related mechanism for wave function collapse, formulated within a Newtonian regime, in which the gravitational self energy associated with spatial superpositions sets the collapse rate. These models provide mechanisms for the suppression of macroscopic quantum superpositions while reproducing standard quantum mechanical predictions at microscopic scales.
   
   Furthermore, a wide range of experimental efforts has been devoted to testing and constraining collapse models \cite{Donadi:2020kzc,Piscicchia:2024wan,deSouza:2024ymq,Tagg:2024fvq,DiBartolomeo:2024wav,Proceedings:2023mkp,ICECUBE:2023gdv,Fadel:2023ici,Napolitano:2023lar,Bassi:2022ibq,Simonov:2022epo,Polkovnikov:2022hkg,Curceanu:2022yeg,Oppenheim:2022xjr,Gasbarri:2021sdm,Belenchia:2021rfb,Carlesso:2016khv}; see also Ref.~\cite{Carlesso:2022pqr} for a comprehensive review. These tests aim to validate or place bounds on the parameters of models such as CSL and DP. In particular, noninterferometric experiments probe indirect signatures of collapse dynamics, including Brownian like motion induced by stochastic localization processes and spontaneous radiation arising from fluctuations in mass or charge density. Constraints on CSL parameters, such as the collapse rate $\lambda$ and correlation length $r_C$, have been obtained from heating rates in Bose-Einstein condensates, position noise in mechanical oscillators, and searches for spontaneous X ray or gamma ray emission. For instance, for $r_C \sim 10^{-7}\,\mathrm{m}$, the collapse rate is constrained to $\lambda \lesssim 10^{-12}\,\mathrm{s}^{-1}$, with weaker bounds at larger correlation lengths. Gravitational wave detectors have also been employed to set bounds on collapse models. In particular, the analysis of noise spectra in detectors such as LIGO and Virgo has been used to constrain both DP and CSL scenarios \cite{Carlesso:2016khv}. These studies place stringent upper limits on collapse induced noise, further restricting the viable parameter space of phenomenological collapse models.
   
   Despite these advances, a key open question remains that whether gravity related decoherence can be derived from an underlying microscopic quantum field theory framework, rather than introduced phenomenologically, and how the resulting dynamics compares with the predictions of collapse models. Several approaches have investigated gravitationally induced decoherence using effective field theory, open quantum systems, and master equation techniques \cite{Blencowe2013,AnastopoulosHu2013,OnigaWang2017}. In particular, graviton emission and gravitational bremsstrahlung have been considered as possible mechanisms through which the gravitational field acquires information about spatially separated matter configurations, thereby reducing their quantum coherence \cite{Riedel2013}. More recently, graviton induced which path decoherence in matter wave interferometry has been investigated by explicitly accounting for the radiative gravitational field generated by different branches of a spatial superposition \cite{Matsui:2026}. Other studies have considered decoherence arising from quantum fluctuations of the gravitational field in different regimes. These include dephasing induced by a background bath of long wavelength gravitons \cite{DeLorenciFord:2015}, noise and decoherence induced by gravitons in the vacuum state \cite{Kanno:2021noise}, and the loss or protection of coherence in a quantum harmonic oscillator coupled to a graviton bath \cite{Toros:2024coherence}. The influence of gravitational wave bursts on spatial superpositions has also been investigated, with particular attention to the contributions of gravitational wave memory and oscillatory radiation \cite{LintonTiwari:2025}. Although these approaches address related aspects of gravitationally induced decoherence, they involve different physical settings and mechanisms, ranging from quantum fluctuations of the gravitational field to radiative graviton emission and interactions with classical gravitational wave backgrounds. Their predictions therefore need not coincide, and their comparison requires careful consideration of the underlying interactions, environmental states, and approximations.
   
   In this context, the QBE provides a natural and powerful tool. Originally developed to describe neutrino flavor evolution in media \cite{Sigl:1993ctk} and the polarization dynamics of cosmic microwave background photons \cite{Kosowsky:1994cy,Bavarsad:2009hm,Bartolo:2018igk,Bartolo:2019eac,Hoseinpour:2020hic}, the QBE offers a momentum resolved description of quantum systems interacting with an environment. Formulated within the Born-Markov approximation, the QBE assumes weak system–environment coupling and negligible memory effects, allowing for a controlled and systematic incorporation of interactions derived directly from quantum field theory \cite{Zarei:2021dpb}. Its collision term describes the effects of emission and absorption processes on both the populations and coherences of a quantum system. This makes it particularly well suited for investigating decoherence processes induced by the emission and absorption of gravitational quanta.
   
   Our previous work has explored complementary aspects of this problem. In Ref.~\cite{Sharifian2024}, we investigated the gravitational decoherence of spin-\(1/2\) particles within an open quantum system framework, considering their interaction with squeezed gravitational waves. In a different context, we developed a QBE approach to describe decoherence through bremsstrahlung emission in the interaction between fermions and photons \cite{ManshouriZarei2025}. This work demonstrates how the collision term of the QBE can be employed to describe the loss of coherence associated with radiative processes and to compare the resulting dephasing with classical bremsstrahlung. These studies motivate extending the same microscopic framework to the interaction between fermions and gravitons.
   
   In this work, we consider a two level quantum system represented by a fermionic spinor field prepared in a spatial superposition. We first analyze graviton emission from this system through a gravitational bremsstrahlung process. Using the collision term of the QBE, we show that this process induces decoherence in the quantum state. The corresponding decoherence rate is derived as a function of the spatial separation of the superposed branches, the gravitational coupling strength, and the particle mass. Our result contains an additional Gaussian factor in its spatial separation dependence, originating from the finite spatial width of the interaction profile. This factor modifies the functional form of the decoherence rate and introduces a characteristic length scale into its spatial dependence. We examine this feature and compare the resulting decoherence law with the predictions of collapse models, particularly the CSL and DP frameworks. Rather than identifying the derived dynamics with a specific collapse model, we investigate the similarities and differences in their predicted suppression of spatial coherence. This comparison allows us to explore how microscopic decoherence due to graviton bremsstrahlung may contribute to our understanding of the physical mechanisms underlying collapse phenomenology.


The remainder of this paper is organized as follows.
In Sec.~\ref{sec:spin-graviton} we review the spin-$1/2$ spinor formalism
and the relevant graviton degrees of freedom, including an overview of the
QBE framework.
Section~\ref{sec:bremsstrahlung} derives the decoherence rate from graviton
bremsstrahlung emission, with examining its limiting
behavior and comparing the result with CSL and
DP collapse models.
Numerical estimates and the emergence of classicality are presented in
Sec.~\ref{sec:numerical}, where we discuss coherent amplification and the
hierarchy of physical regimes.
Section~\ref{sec:conclusion} summarizes our findings and outlines
directions for future work.
Technical details of the derivations are collected in the Appendix.

 \section{Spin-$1/2$ and Graviton Degrees of Freedom} 
 \label{sec:spin-graviton} 
 In the following, we investigate the graviton emission from a qubit system parametrized by a spin-$1/2$ spinor field in spatial superposition. To analyze the interaction between spin-$1/2$ particles and gravitons, we begin by quantizing the spinor and graviton fields. 
 We decompose the spinor field into its creation and annihilation parts, denoted by $\psi^+(x)$ and $\psi^-(x)$, and expand these parts as
 \begin{eqnarray} \label{psi1}
 	\psi(x) = \psi^+(x) + \psi^-(x)~,
 \end{eqnarray}
 where
 \begin{eqnarray} \label{psiplus}
 	\psi^+(x) = \int d\q \sum_{r} u_r(\q) \hat{c}_r(\q) e^{-i(q^0 t - \mathbf{q} \cdot \mathbf{x})}~, \\
 	\label{psi2}
 	\psi^-(x) = \int d\q \sum_{r} \bar{u}_r(\q) \hat{c}^\dag_r(\q) e^{i(q^0 t - \mathbf{q} \cdot \mathbf{x})}~,
 \end{eqnarray}
 in which the spin is labeled by $r=1,2$, and $u_r$ represents the nonrelativistic free particle spinor. In the nonrelativistic regime relevant for the localized qubit systems considered here, the Dirac spinor reduces to
 \begin{equation}
 	u_{s}(\q)
 	\simeq
 	\begin{pmatrix}
 		\chi_{s} \\
 		\dfrac{\boldsymbol{\sigma}\cdot \q}{2m_f}\chi_{s}
 	\end{pmatrix}~,
 \end{equation}
 where $m_f$ is the fermion mass, \(\chi_s\) are two component Pauli spinors defined as
 \begin{equation}
 	\chi_1 =
 	\begin{pmatrix}
 		1 \\
 		0
 	\end{pmatrix}~,
 	\qquad
 	\chi_2 =
 	\begin{pmatrix}
 		0 \\
 		1
 	\end{pmatrix}~.
 \end{equation}
 Furthermore, there exists an anticommutation relation between the creation and annihilation operators
 \begin{align}
 	\left\{ \hat{c}_r(\mathbf{q}), \hat{c}^\dag_{r'}(\mathbf{q'}) \right\} 
 	&= (2\pi)^3 \delta^3(\mathbf{q} - \mathbf{q'}) \delta_{r r'}~.
 \end{align}
 One can describe the density operator of a system of neutral atoms in the following form
 \begin{equation}\label{dno}
 	\hat{\rho}^{(f)} = \int d\q' \rho^{(f)}_{ij}(\mathbf{q'}) \hat{c}_i^\dag(\mathbf{q'}) \hat{c}_j(\mathbf{q'})~,
 \end{equation}
 where the macroscopic properties of the spin-$1/2$ particle in the interferometer are encoded in the density matrix $\rho^{(f)}_{ij}$. The expectation value of the spin-$1/2$ number operator can be expressed as
 \begin{equation}\label{efn}
 	\left\langle \mathcal{\hat{D}}_{ij}(\mathbf{q}) \right\rangle = \text{Tr} \left( \hat{\rho}^{(f)}\, \mathcal{\hat{D}}_{ij} \right) = (2\pi)^3 \delta^3(0) \rho^{(f)}_{ji}(\mathbf{q})~,
 \end{equation}
 where $\rho^{(f)}_{ij}(\mathbf{q})$ is the density matrix of a system of spin-$1/2$ particles. 
 On the other hand, when quantum fluctuations of the metric tensor embedding the gravitational interaction are negligible, any purported theory of quantum gravity should reduce to quantum field theory in curved spacetime. 
 We now consider the soft graviton degrees of freedom, which are affected by their coupling to the environment. The tensor field $h_{\mu\nu}$ is given by assuming the weak field limit and expanding the metric around Minkowski spacetime as follows
 \begin{equation}  
 	g_{\mu\nu} = \eta_{\mu\nu} + \kappa h_{\mu\nu}~,
 \end{equation}
 where $\kappa = \sqrt{16\pi G}$ is the gravitational coupling constant. The associated quantum field is decomposed as
 \begin{equation}\label{quant}
 	h_{\mu\nu}(x) = h^+_{\mu\nu}(x) + h^-_{\mu\nu}(x)~,
 \end{equation}
 where $h^-_{\mu\nu}(x)$ and $h^+_{\mu\nu}(x)$ are linear in graviton creation and annihilation operators, respectively. The Fourier transforms of the fields are given by
 \begin{eqnarray} \label{h1}
 	h^+_{\mu\nu}(x) &= \int d\mathbf{p} \sum_{s = +, \times} a_s(\mathbf{p}) h^s_{\mu\nu}(p) e^{-i(p^0 t - \mathbf{p} \cdot \mathbf{x})}~, \\ \label{h2}
 	h^-_{\mu\nu}(x) &= \int d\mathbf{p} \sum_{s = +, \times} a^\dag_{s}(\mathbf{p}) h^{s \ast}_{\mu\nu}(p) e^{i(p^0 t - \mathbf{p} \cdot \mathbf{x})}~,
 \end{eqnarray}
 where $a_s(\mathbf{p})$ and $a_s^\dag(\mathbf{p})$ are the graviton annihilation and creation operators that obey the commutation relation
 \begin{equation} \label{commutationa}
 	\left[ a_s(\mathbf{p}), a^\dag_{s'}(\mathbf{p'}) \right] = (2\pi)^3 2p^0 \delta^3(\mathbf{p} - \mathbf{p'}) \delta_{ss'}~.
 \end{equation}
 The polarization tensors $h^{(s)}_{\mu\nu}$ have the following well known properties
 \begin{eqnarray}
 	h^s_{\mu\nu}(p) p^\mu = 0~, \quad h_\mu^{\mu}(p) = 0~, \quad
 	h^s_{\mu\nu}(p) \left(h^{s'}_{\mu\nu}(p)\right)^{\ast} = \delta^{ss'}~ \label{canonical}.
 \end{eqnarray}
 Note that in equations \eqref{h1} and \eqref{h2}, we have not separated the microscopic and mesoscopic times to avoid confusion. We assume that the time appearing in the exponential function is a microscopic time, while the annihilation and creation operators can generally be a function of both times. 
 It is also convenient to represent the polarization tensor $h^{(s)}_{\mu\nu}$ in terms of a direct product of unit spin polarization vectors
 \begin{eqnarray}
 	h^s_{\mu\nu}(p) = e^s_\mu(p) e^s_\nu(p)~, \quad e^s_\mu(p) p^\mu = 0~, \quad
 	\left[ e^s_\mu(p) \left(e^{s'\mu}(p)\right)^\ast \right]^2 = \delta^{ss'}~.
 \end{eqnarray}
 The graviton density operator is presented in the following form \cite{Bartolo:2018igk}
 \begin{equation} \label{rhohatg}
 	\hat{\rho}^{(\textrm{g})} = \int \frac{d^3p}{(2\pi)^3} \rho^{(\textrm{g})}_{ij} a^\dag_i(\mathbf{p}) a_j(\mathbf{p})~,
 \end{equation}
 where, for the unpolarized graviton, the density matrix is given by
 \begin{equation}
 	\rho^{(\textrm{g})} = \frac{1}{2} \begin{pmatrix} I^{(\textrm{g})} & 0 \\ 0 & I^{(\textrm{g})} \end{pmatrix}~, \label{Ig}
 \end{equation}
 where $I^{(\textrm{g})}$ denotes the radiation intensity.

 \subsection{Overview of QBE } 
 \label{QBE}
 As mentioned we investigate the emission of gravitons from a fermionic system in a spatial superposition state, examining the impact of this radiation on the system's coherence. This interaction will be studied microscopically using field theory methods. To model the graviton emission from the fermionic system, we use the QBE to describe the time evolution of the associated density matrix  \cite{Kosowsky:1994cy,Bavarsad:2009hm,Bartolo:2018igk,Bartolo:2019eac,Hoseinpour:2020hic}
 \bea
 &&(2\pi)^3\delta^3(0)\dot{\rho}^{(f)}_{ij}(\k,t)=i\left\langle\left[H_\inte,\hat{\mathcal{D}}_{ij}(\k)\right]\right \rangle
 +
 D_{ij}[\rho^{(f)}]~,\label{QBE1}
 \eea
 with
 \bea
 D_{ij}[\rho^{(f)}]=-
 \int_{0}^{t}ds\left<\l[H_\inte(s),
 \l[H^\dagger_\inte(0),\hat{\mathcal{D}}_{ij}(\k)\r]\r]\right>~, \label{QBE1-1}
 \eea 
 where the interaction Hamiltonian is defined in terms of the 
 $S$-matrix elements as
 \beq 
 S=-i\int_{-\infty}^{\infty}dt H_\inte(t)~.
 \eeq 
 It is crucial to note that $H_\inte(t)$ defined using the $S$-matrix elements will describe a specific graviton emission process. 
 Note that in this expression, an imaginary factor 
 $i$ has been factored out.
 Also $\rho^{(f)}_{ij}$ is the density matrix of the fermion system,
 $\hat{\mathcal{D}}_{ij}(\k)=c^\dag_i(\k) c_j(\k)$ is the number operator of the fermionic system and the expectation value $\left < \cdot\cdot\cdot\right >$ of a generic operator $A$ is given by
 \cite{Kosowsky:1994cy,Bavarsad:2009hm,Bartolo:2018igk,Bartolo:2019eac,Hoseinpour:2020hic}
 \beq \label{exp1}
 \left < A(\k)\right >=\mathrm{tr}[\hat\rho ^{(f)}A(\k)]=\int d\q\left<\q|\hat{\rho}^{(f)} A(\k)|\q \right >~,
 \eeq
 with the abbreviation $d\q=d^3q/(2\pi)^3$ and $\hat{\rho}$ denotes the following density operator 
 \beq
 \hat{\rho}^{(f)}=\int d\q\,\rho^{(f)}_{ij}(\q)\hat{\mathcal{D}}_{ij}(\q)~.
 \eeq
 In QBE \eqref{QBE1} the first term on the right hand side represents the forward scattering contribution responsible for phase generation affecting quantum coherence.
 The second term on the right side is the dissipator term responsible for decoherence and related dissipative effects.

 \begin{figure}[tb]
 	\centering
 	\includegraphics[scale=0.4]{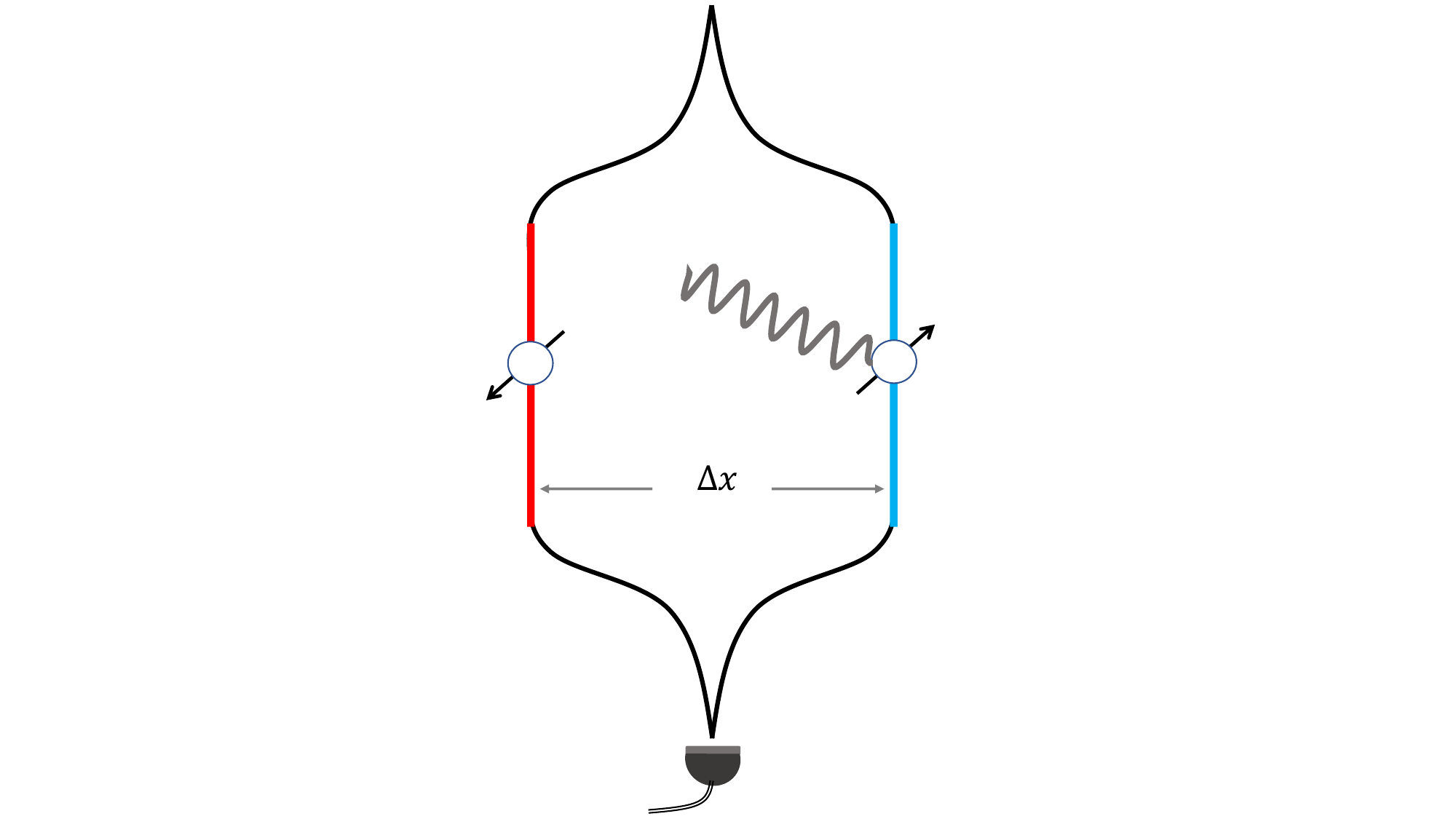}
 	\caption{Graviton bremsstrahlung process from a fermion in a spatial superposition state.}
 	\label{Bre1}
 \end{figure}

 \section{Interaction of spin-$1/2$ systems with graviton: bremsstrahlung emission}
 \label{sec:bremsstrahlung}
 
 In the QBE, the interaction Hamiltonian is described in terms of quantum fields.
 Specifically, the qubit-graviton interaction is modeled as the coupling between a
 spinor field and a tensor field. As mentioned, the spacetime fluctuation $h_{\mu\nu}$
 is assumed to be small, allowing for the linear approximation of general relativity
 to hold. In the nonrelativistic limit, the interaction with the fermion system
 simplifies to involve only the spin operator. This graviton bremsstrahlung emission
 process is described by the following effective interaction Hamiltonian
 \cite{Zarei:2021dpb,Sharifian:2023jem,Zarei:2025nnm}
 \begin{equation}
 	H_{\mathrm{int}}(t)
 	=
 	-i\frac{\kappa}{2}
 	\int d^3x\,d\tau\,
 	h^{-\mu\nu}(x)\,
 	\partial_{\mu}\bar\psi^-(x)\,
 	\gamma_{\nu}\,
 	\psi^+(x)\,
 	V\delta_{\sigma_0}^4(x-\bar x(\tau))~,
 	\label{Hintx}
 \end{equation}
 where $\kappa^2=16\pi G$ and the factor $\delta_{\sigma_0}^4(x-\bar x(\tau))$ is due
 to the spatial localization of the matter degrees of freedom around a classical
 worldline $\bar x(\tau)$ with $\tau$ denoting the proper time
 \cite{Breuer:2001,Sharifian:2025olk}. One can model this distribution using a
 Gaussian wavepacket profile multiplied by a time dependent Dirac delta function
 \cite{Breuer:2001,Sharifian:2025olk}
 \begin{equation}
 	\delta_{\sigma_0}^4(x-\bar x(\tau))
 	=
 	\delta(x^0-\bar x^0(\tau))
 	\frac{1}{(2\pi\sigma_0^2)^{3/2}}
 	\exp\!\left[
 	-\frac{|\mathbf x-\bar{\mathbf x}(\tau)|^2}{2\sigma_0^2}
 	\right]~.
 \end{equation}
 This expression describes the finite spatial extension of the localized matter system
 with width $\sigma_0$.
 
 Next, we insert the Fourier transforms \eqref{psi1}, \eqref{psi2}, and \eqref{h1}
 into \eqref{Hintx} and obtain
 \begin{equation}
 	\begin{aligned}
 		H_{\mathrm{int}}
 		={}&
 		\frac{\kappa}{2}
 		\sum_{s,r,r'}
 		\int d^3x\,d\tau\,d\p\,d\q\,d\q'\,
 		h^{s\mu\nu}(\p)\,
 		q'_\mu\,
 		\bar u_{r'}(\q')\,
 		\gamma_\nu\,
 		u_r(\q)\,
 		V\delta_{\sigma_0}^4(x-\bar x(\tau))\,
 		e^{i(q'-q+p)\cdot x}\,
 		a_s(\p)\,
 		c^\dagger_{r'}(\q')\,
 		c_r(\q)~.
 	\end{aligned}
 	\label{hgnu2}
 \end{equation}
 After integrating over $\mathbf x$ we get
 \begin{equation}
 	\begin{aligned}
 		H_{\mathrm{int}}
 		={}&
 		\frac{\kappa}{2}V
 		\sum_{s,r,r'}
 		\int d\tau\,d\p\,d\q\,d\q'\,
 		h^{s\mu\nu}(\p)\,
 		\bar u_{r'}(\q')\,
 		(iq'_\mu)\,
 		\gamma_\nu\,
 		u_r(\q)\,
 		e^{-|\q'-\q-\p|^2\sigma_0^2/2}\,
 		e^{-i(\q'-\q-\p)\cdot\bar{\mathbf x}(\tau)}\,
 		e^{i(q'^0-q^0-p^0)x^0}
 		\\
 		&\qquad\times
 		\delta(x^0-\bar x^0(\tau))\,
 		a_s(\p)\,
 		c^\dagger_{r'}(\q')\,
 		c_r(\q)~.
 	\end{aligned}
 	\label{Hint-2}
 \end{equation}
 We then insert the interaction Hamiltonian \eqref{Hint-2} into the dissipator term
 of the QBE \eqref{QBE1}; the details are given in Appendix~A.
 
 Before tracing, the joint state of the fermion (path $\oplus$ spin) and emitted
 graviton takes the entangled form
 \begin{equation}
 	|\Psi\rangle
 	=
 	\frac{1}{\sqrt{2}}
 	\Bigl(
 	|1,{\uparrow}\rangle|g_1\rangle
 	+|2,{\downarrow}\rangle|g_2\rangle
 	+\cdots
 	\Bigr),
 \end{equation}
 where $|s,\sigma\rangle|g_s\rangle$ denotes branch $s$ with spin $\sigma$ and the
 corresponding graviton state. Tracing over both spin and graviton degrees of freedom
 yields the reduced spatial density matrix whose off diagonal element is $\rho_{12}$.
 
 We focus on the off diagonal coherence element $\rho_{12}(t)$, whose decay encodes
 environment induced loss of quantum phase information. From the fully reduced
 gravitational master equation, all spin, polarization, and Lorentz indices have been
 traced out, yielding a closed evolution equation of the form
 \begin{equation}
 	\dot\rho_{12}(t)
 	=
 	-\Gamma(\Delta\mathbf{x})\,\rho_{12}(t)~,
 \end{equation}
 which describes a pure dephasing process.
 
 The decoherence rate is given by
 \begin{equation}
 	\Gamma(\Delta\mathbf{x})
 	=
 	\frac{\kappa^2}{16m_f^2}\,V
 	\int_0^\infty\frac{p^2\,dp}{2\pi^2}\,
 	\tilde{I}^{(g)}(p)\,
 	\left[
 	\mathcal{G}(\Delta\mathbf{x})
 	-
 	\frac{1}{4\pi}
 	\int d\Omega_{\hat{\mathbf{p}}}\,
 	\mathcal{F}(\hat{\mathbf{p}})\,
 	\mathcal{J}_{\sigma_p}(\mathbf{p},\Delta\mathbf{x})
 	\right]~,
 	\label{GammaFull}
 \end{equation}
 where the graviton spectral density is defined as
 \begin{equation}
 	\tilde{I}^{(g)}(p)
 	=
 	\int d\tau\,e^{-ip\tau}I^{(g)}(p)~,
 \end{equation}
 and the total graviton number in the normalization volume is
 \begin{equation}
 	N_g
 	=
 	V\int_0^\infty\frac{p^2\,dp}{2\pi^2}\,\tilde{I}^{(g)}(p)~.
 \end{equation}
 The spatial overlap kernel $\mathcal{G}(\Delta\mathbf{x})$ is given in
 Eq.~\eqref{mathG},
 \begin{equation}
 	\mathcal{G}(\Delta\mathbf{x})
 	=
 	\frac{1}{(8\pi\sigma_0^2)^{3/2}}
 	\exp\!\left(
 	-\frac{|\Delta\mathbf{x}|^2}{8\sigma_0^2}
 	\right)~,
 	\label{mathG}
 \end{equation}
 and encodes the finite spatial width $\sigma_0$ of the localized branches. The
 angular factor $\mathcal{F}(\hat{\mathbf{p}})$ is the TT polarization factor obtained
 after summing over physical graviton polarizations,
 \begin{equation}
 	\mathcal{F}(\theta,\phi)
 	=
 	\sin^2\theta + \cos^2\theta\sin^2\phi~,
 	\qquad
 	\int d\Omega_{\hat{\mathbf{p}}}\,\mathcal{F}(\hat{\mathbf{p}})
 	=
 	\frac{10\pi}{3}~.
 \end{equation}
 The kernel $\mathcal{J}_{\sigma_p}(\mathbf{p},\Delta\mathbf{x})$ is the
 $\mathbf{q}$ integrated fermionic momentum distribution. We do not approximate
 $\rho^{(f)}_{r''r'}(\mathbf{q})$ by its value at $\mathbf{q}=\mathbf{k}$; instead
 we use the Gaussian ansatz
 \begin{equation}
 	\rho^{(f)}_{r''r'}(\mathbf{q})
 	=
 	\rho^{(f)}_{r''r'}(\mathbf{k})
 	\exp\!\left[
 	-\frac{|\mathbf{q}-\mathbf{k}|^2}{2\sigma_p^2}
 	\right]~,
 	\label{Gaussianrho}
 \end{equation}
 with momentum space width $\sigma_p$. With this profile,
 \begin{equation}
 	\begin{aligned}
 		\mathcal{J}_{\sigma_p}(\mathbf{p},\Delta\mathbf{x})
 		={}&
 		\int\frac{d^3\mathbf{q}}{(2\pi)^3}
 		\exp\!\left[-2\sigma_0^2|\mathbf{k}-\mathbf{q}+\mathbf{p}|^2\right]
 		\exp\!\left[-i(\mathbf{k}-\mathbf{q}+\mathbf{p})\cdot\Delta\mathbf{x}\right]
 		\exp\!\left[-\frac{|\mathbf{q}-\mathbf{k}|^2}{2\sigma_p^2}\right]
 		\\
 		={}&
 		\frac{1}{8\pi^{3/2}}
 		\left(
 		\frac{2\sigma_p^2}{1+4\sigma_0^2\sigma_p^2}
 		\right)^{\!3/2}
 		\exp\!\left[
 		-\frac{2\sigma_0^2 p^2}{1+4\sigma_0^2\sigma_p^2}
 		-\frac{\sigma_p^2|\Delta\mathbf{x}|^2}{2(1+4\sigma_0^2\sigma_p^2)}
 		-\frac{i\,\mathbf{p}\cdot\Delta\mathbf{x}}{1+4\sigma_0^2\sigma_p^2}
 		\right]~.
 	\end{aligned}
 	\label{calJ}
 \end{equation}
 Equation~\eqref{GammaFull} together with Eq.~\eqref{calJ} is the full decoherence
 factor obtained in this work. In the limit of broad momentum width,
 $\sigma_p\to\infty$, the Gaussian weight in $\mathbf{q}$ becomes uniform, the
 prefactor in Eq.~\eqref{calJ} approaches $(2\sigma_0^2)^{-3/2}/(8\pi^{3/2})$, and
 one recovers
 \begin{equation}
 	\mathcal{J}_{\sigma_p}(\mathbf{p},\Delta\mathbf{x})
 	\;\xrightarrow{\sigma_p\to\infty}\;
 	\mathcal{G}(\Delta\mathbf{x})\,
 	e^{-i\mathbf{p}\cdot\Delta\mathbf{x}}~,
 \end{equation}
 so that the angular integral in Eq.~\eqref{GammaFull} yields
 \begin{equation}
 	\frac{1}{4\pi}\int d\Omega_{\hat{\mathbf{p}}}\,
 	\mathcal{F}(\hat{\mathbf{p}})\,e^{-i\mathbf{p}\cdot\Delta\mathbf{x}}
 	=
 	\frac{5}{6}\,\frac{\sin(p|\Delta\mathbf{x}|)}{p|\Delta\mathbf{x}|}~,
 \end{equation}
 and Eq.~\eqref{GammaFull} reduces to the factorized form
 \begin{equation}
 	\Gamma(\Delta\mathbf{x})
 	=
 	\frac{\kappa^2}{16m_f^2}\,V\,
 	\mathcal{G}(\Delta\mathbf{x})
 	\int_0^\infty\frac{p^2\,dp}{2\pi^2}\,
 	\tilde{I}^{(g)}(p)
 	\left[
 	1
 	-
 	\frac{5}{6}\,\frac{\sin(p|\Delta\mathbf{x}|)}{p|\Delta\mathbf{x}|}
 	\right]~.
 	\label{GammaStandard}
 \end{equation}
 Thus the bracket $1-\tfrac{5}{6}\sin(p|\Delta\mathbf{x}|)/(p|\Delta\mathbf{x}|)$ is
 recovered as the broad momentum, factorized limit of the full kernel. The essential
 new ingredient of Eq.~\eqref{GammaFull} is the Gaussian factor
 $\exp[-\sigma_p^2|\Delta\mathbf{x}|^2/(2(1+4\sigma_0^2\sigma_p^2))]$ inside
 $\mathcal{J}_{\sigma_p}$, which introduces an additional coherence length associated
 with the finite momentum spread of the superposed wavepackets.
 
 The solution of the dephasing equation integrates exactly to
 \begin{equation}
 	\rho_{12}(t)
 	=
 	\rho_{12}(0)\,
 	\exp\!\bigl[-\Gamma(\Delta\mathbf{x})\,t\bigr]~.
 	\label{rho12}
 \end{equation}
 Here the labels $1$ and $2$ denote the two branches of the spatial superposition
 separated by $\Delta\mathbf{x}$; all spin ($i,j$) and graviton polarization indices
 have been traced over in arriving at this scalar equation.

  \subsection{Limiting behavior of the decoherence factor}
  
  In the short-distance limit $|\Delta\mathbf{x}|\to 0$, the angular
  integral in Eq.~\eqref{GammaFull} becomes independent of
  $\hat{\mathbf{p}}$, and
  \begin{equation}
  	\mathcal{J}_{\sigma_p}(\mathbf{p},0)
  	=
  	\frac{1}{8\pi^{3/2}}
  	\left(
  	\frac{2\sigma_p^2}{1+4\sigma_0^2\sigma_p^2}
  	\right)^{3/2}
  	\exp\!\left[
  	-\frac{2\sigma_0^2 p^2}{1+4\sigma_0^2\sigma_p^2}
  	\right]~.
  \end{equation}
  Requiring that the fermionic momentum profile be normalized consistently
  with the spatial kernel at coincidence,
  $\tfrac{1}{4\pi}\int d\Omega_{\hat{\mathbf{p}}}\,
  \mathcal{F}(\hat{\mathbf{p}})\,\mathcal{J}_{\sigma_p}(\mathbf{p},0)
  =\mathcal{G}(0)$ for all $p$ contributing to the $p$-integral, the
  bracket in Eq.~\eqref{GammaFull} vanishes at $\Delta\mathbf{x}=0$,
  giving
  \begin{equation}
  	\Gamma(0)=0~.
  \end{equation}
  This is the expected pure-dephasing result: coincident branches carry no
  which-path information, and no gravitational decoherence occurs.
  
  The large-separation limit requires distinguishing two idealized
  treatments.  If the broad-momentum limit $\sigma_p\to\infty$ is taken
  before $|\Delta\mathbf{x}|\to\infty$, then Eq.~\eqref{GammaFull}
  reduces to Eq.~\eqref{GammaStandard}
  \begin{equation}
  	\Gamma(\Delta\mathbf{x})
  	=
  	\frac{\kappa^2}{16m_f^2}\,V\,
  	\mathcal{G}(\Delta\mathbf{x})
  	\int_0^\infty\frac{p^2\,dp}{2\pi^2}\,\tilde{I}^{(g)}(p)
  	\left[
  	1-\frac{5}{6}\frac{\sin(p|\Delta\mathbf{x}|)}{p|\Delta\mathbf{x}|}
  	\right],
  \end{equation}
  and
  \begin{equation}
  	\frac{\sin(p|\Delta\mathbf{x}|)}{p|\Delta\mathbf{x}|}
  	\to 0~,
  	\qquad
  	|\Delta\mathbf{x}|\to\infty~.
  \end{equation}
  In this regime the sinc bracket saturates at unity, yet the full rate
  \eqref{GammaStandard} still carries the prefactor
  $\mathcal{G}(\Delta\mathbf{x})\propto
  e^{-|\Delta\mathbf{x}|^2/8\sigma_0^2}$, which suppresses $\Gamma$
  whenever the branch separation exceeds the wavepacket width $\sigma_0$.
  This Gaussian envelope has a direct physical origin: a graviton whose
  spatial coherence length is $\sim\sigma_0$ cannot resolve two branches
  separated by $|\Delta\mathbf{x}|\gg\sigma_0$, so the emitted radiation
  carries no which-path information and the decoherence rate vanishes.
  A separation-independent rate therefore requires the additional
  idealization $\sigma_0\to\infty$, in which $\mathcal{G}$ is renormalized
  to a constant reference value and the localized-branch structure is
  washed out.
  
  Under the additional idealization $\sigma_0\to\infty$ (on top of the
  $\sigma_p\to\infty$ limit that gave \eqref{GammaStandard}), the
  spatial prefactor
  $\mathcal{G}(\Delta\mathbf{x})\propto e^{-|\Delta\mathbf{x}|^2/8\sigma_0^2}$
  tends to a constant, and one obtains the maximal rate
  \begin{equation}
  	\Gamma(\infty)
  	=
  	\frac{\kappa^2}{16m_f^2}\,V
  	\int_0^\infty\frac{p^2\,dp}{2\pi^2}\,\tilde{I}^{(g)}(p)
  	=
  	\frac{\kappa^2}{16m_f^2}\,N_g~.
  	\label{Gammamax}
  \end{equation}
  This is a formal upper bound: it requires both the fermionic momentum
  distribution and the matter wavepacket to be taken to their
  (non-normalizable) plane-wave extremes, and it is not attained for any
  state with finite $\sigma_0$ or $\sigma_p$.
  
  For finite $\sigma_p$ and $\sigma_0$, the full rate
  \eqref{GammaStandard} carries the prefactor
  $\mathcal{G}(\Delta\mathbf{x})\propto e^{-|\Delta\mathbf{x}|^2/8\sigma_0^2}$
  together with the factor
  \begin{equation}
  	\exp\!\left[
  	-\frac{\sigma_p^2\,|\Delta\mathbf{x}|^2}{2(1+4\sigma_0^2\sigma_p^2)}
  	\right]
  \end{equation}
  inside $\mathcal{J}_{\sigma_p}$.  Both suppress the rate exponentially
  once $|\Delta\mathbf{x}|$ exceeds the relevant wavepacket scales, so
  \begin{equation}
  	\Gamma(\Delta\mathbf{x})\to 0
  	\qquad
  	(|\Delta\mathbf{x}|\to\infty)
  \end{equation}
  for any fixed finite $\sigma_0$ and $\sigma_p$.
  
  The physical origin of this suppression is straightforward.  A graviton
  whose spatial coherence length is set by $\sigma_0$ and $\sim 1/\sigma_p$
  cannot resolve two branches separated by $|\Delta\mathbf{x}|$ much
  larger than those scales: the bremsstrahlung emitted from either branch
  is essentially identical, it carries no which-path information, and the
  decoherence rate vanishes.  The vanishing of $\Gamma$ at large
  separation is therefore a consequence of the finite phase-space support
  of the superposed matter state, not a failure of the emission mechanism.
  For any normalizable wavepacket the rate vanishes at
  $|\Delta\mathbf{x}|=0$ (the two branches coincide and there is no path
  distinguishability), rises to a maximum at intermediate separations of
  order $\sigma_0$ and $1/\sigma_p$, and then falls back to zero via the
  combined Gaussian envelope.  The idealized value \eqref{Gammamax}
  therefore serves as a formal reference scale rather than an attainable
  rate.

  \subsection{Comparison with other gravitational decoherence mechanisms}
  
  The behavior obtained above should be contrasted with the predictions of
  collapse models and with other gravitational decoherence channels.
  In the CSL model
  \cite{Smirne:2014paa,Bassi:2012bg}, the decoherence rate grows as
  $\Delta x^2$ for separations small compared with the correlation length
  $r_C$, but saturates for $\Delta x\gg r_C$ at a maximum value that scales
  quadratically with the mass,
  $\Gamma_{\rm CSL}^{\max}=\lambda_{\rm CSL}(m/m_0)^2$.
  The DP model \cite{Diosi:1989hlx,Penrose:1996cv},
  by contrast, predicts a rate of the form
  \begin{equation}
  	\Gamma_{\rm DP}(\Delta x)
  	=
  	\frac{Gm^2}{\hbar\sqrt{\pi}R_0}
  	\left(
  	1-\frac{R_0}{\sqrt{R_0^2+\Delta x^2}}
  	\right)~,
  \end{equation}
  which also saturates at large separations, approaching the finite ceiling
  $Gm^2/(\hbar\sqrt{\pi}R_0)$ from below with power law corrections of
  order $R_0/|\Delta\mathbf{x}|$.
  Neither of these phenomenological gravity based collapse models predicts a
  strictly vanishing rate at $|\Delta\mathbf{x}|\to\infty$ for a pointlike
  mass distribution: the DP rate saturates from below with a slow power law
  approach, while the CSL rate approaches a constant floor directly.
  
  The bremsstrahlung channel considered here is a radiative, emission based
  mechanism, and its large distance suppression is tied to the finite overlap
  of the superposed wavepackets.
  Master equation treatments of linearized gravity
  \cite{Blencowe2013,AnastopoulosHu2013} have emphasized that decoherence
  due to graviton scattering appears only at order $G^2$ (or higher), whereas
  the emission/bremsstrahlung contribution can vanish in the low frequency
  limit.
  Related studies of decoherence induced by long wavelength gravitons
  \cite{DeLorenciFord:2015}, graviton noise and decoherence
  \cite{Kanno:2021noise}, loss of coherence from a graviton bath
  \cite{Toros:2024coherence}, decoherence by gravitational waves
  \cite{LintonTiwari:2025}, and graviton induced which path decoherence in
  matter wave interferometry \cite{Matsui:2026} address complementary aspects
  of the same general problem.
  Earlier work on graviton bremsstrahlung as a decoherence channel
  \cite{Riedel2013,ManshouriZarei2025} also stressed the role of emission
  amplitudes and which path information.
  The vanishing of $\Gamma$ at large $|\Delta\mathbf{x}|$ found here is
  therefore consistent with the general picture that radiative gravitational
  decoherence is controlled by the emission amplitude and the resulting
  which path information, rather than by a static gravitational self energy
  that would persist at arbitrarily large separation.
  It also implies that experimental searches for graviton bremsstrahlung
  decoherence should focus on separations comparable to or smaller than the
  wavepacket coherence length, where the full kernel~\eqref{GammaFull}
  differs appreciably from both the saturated CSL prediction and the
  power law approached DP saturation.
  
  The suppression of off diagonal terms implies that the reduced density
  matrix becomes approximately diagonal in the position basis
  \begin{equation}
  	\rho(x,x',t)\to0
  	\qquad
  	\text{for }x\neq x'~,
  \end{equation}
  on timescales governed by $\Gamma(\Delta\mathbf{x})$.
  As a result, gravitationally induced decoherence selects localized
  wavepackets as effective pointer states, leading to emergent classical
  behavior.
  Graviton emission acts as a continuous environmental monitoring process.
  Each emitted graviton carries phase information about the emitting matter
  configuration, effectively encoding which path information into the
  gravitational field.
  Tracing over the graviton degrees of freedom results in irreversible loss
  of phase coherence.

   \section{Emergence of classicality from graviton emission: numerical estimates}
   \label{sec:numerical}

   We begin with the pure dephasing evolution equation Eq.~\eqref{rho12}. This structure
   already implies that graviton emission acts as an environmental monitoring channel,
   suppressing phase coherence without affecting populations.
   
   \subsubsection{Coherent amplification and volume normalized spectrum}
   
   For a composite system of $N$ approximately identical constituents, graviton emission
   amplitudes add coherently in the long wavelength regime. The total emission amplitude
   can be estimated in terms of the amplitude of the emission from a single particle
   $\mathcal{A}_1$ as
   \begin{equation}
   	\mathcal{A}_{\mathrm{tot}}
   	=
   	\sum_{i=1}^{N}\mathcal{A}_i
   	\approx
   	N\,\mathcal{A}_1~,
   \end{equation}
   which leads to a quadratic enhancement at the level of probabilities and therefore
   decoherence rates,
   \begin{equation}
   	\Gamma_N
   	=
   	N^2\,\Gamma_1~,
   	\label{GammaN}
   \end{equation}
   where $\Gamma_1$ is the single particle rate defined in Eq.~\eqref{GammaFull}. 
  In the broad momentum, factorized limit $\sigma_p\to\infty$, this reduces to
  Eq.~\eqref{GammaStandard}; the finite width kernel in Eq.~\eqref{GammaFull}
  adds a Gaussian suppression at separations exceeding the wavepacket coherence
  length $\ell_c = (1 + 4\sigma_0^2\sigma_p^2)^{1/2}/\sigma_p$.
The corresponding evolution equation becomes
   \begin{equation}
   	\dot{\rho}_{12}^{(N)}(t)
   	=
   	-\,\Gamma_N\,\rho_{12}^{(N)}(t)~,
   \end{equation}
   with solution
   \begin{equation}
   	\rho_{12}^{(N)}(t)
   	=
   	\rho_{12}^{(N)}(0)\,
   	\exp\!\left[-N^2\Gamma_1 t\right]~.
   \end{equation}
   The associated decoherence time is therefore
   \begin{equation}
   	\tau_{\mathrm{dec}}^{(N)}
   	=
   	\frac{1}{N^2\Gamma_1}~.
   \end{equation}
   
   For a system of total mass $M=Nm_f$, the scaling can be written as
   \begin{equation}
   	\Gamma_M
   	=
   	\left(\frac{M}{m_f}\right)^2\Gamma_1~,
   \end{equation}
   demonstrating quadratic growth of decoherence with total mass in the coherent regime.

   In the idealized factorized/plane-wave limit, the maximum decoherence rate for an
   electron ($m_f=m_e$) is obtained by taking $\sigma_p\to\infty$ (dropping the sinc
   term) and $\sigma_0\to\infty$ (setting $\mathcal{G}(\Delta\mathbf{x})\to1$), i.e.\
   the same double idealization underlying Eq.~\eqref{Gammamax}
   \begin{equation}
   	\Gamma_e^{\max}
   	\sim
   	\frac{\kappa^2}{16m_e^2}
   	V\int_0^\infty\frac{p^2\,dp}{2\pi^2}\,\tilde{I}^{(g)}(p)
   	=
   	\frac{\kappa^2}{16m_e^2}\,N_g~,
   \end{equation}
   which shows explicitly that the rate is proportional to the total graviton number
   $N_g$. As discussed above, this is a formal upper bound rather than an attainable
   rate: in the full expression Eq.~\eqref{GammaFull}, the factor
   $\mathcal{G}(\Delta\mathbf{x})$ and the Gaussian kernel $\mathcal{J}_{\sigma_p}$
   replace this maximal estimate by a separation-dependent rate, which is suppressed
   once $|\Delta\mathbf{x}|$ exceeds the wavepacket coherence length set by
   $\sigma_0$ and $1/\sigma_p$. Despite this idealization, the Planck-scale
   suppression $\kappa^2\sim G$ ensures that $\Gamma_e^{\max}$ remains extremely
   small for any realistic graviton background.

   \begin{table}[t]
   	\centering
   	\begin{tabular}{cccc}
   		\hline\hline
   		System & Mass scale (kg) & $N$ (nucleons) & Decoherence regime \\
   		\hline
   		Electron           & $10^{-31}$               & 1 ($m_f=m_e$)       & Fully coherent    \\
   		Atom               & $10^{-27}$               & $\sim1$             & Fully coherent    \\
   		Large molecule     & $10^{-24}$--$10^{-21}$   & $10^3$--$10^5$      & Weak decoherence  \\
   		Nanoparticle       & $10^{-18}$--$10^{-15}$   & $10^9$--$10^{12}$   & Transition regime \\
   		Dust grain         & $10^{-12}$               & $\sim10^{15}$       & Rapid decoherence \\
   		Macroscopic object & $\gtrsim10^{-6}$         & $\gtrsim10^{21}$    & Classical limit   \\
   		\hline\hline
   	\end{tabular}
   	\caption{%
   		Graviton induced decoherence across physical systems using mode occupation
   		normalization and the maximal factorized kernel of Eq.~\eqref{GammaStandard}.
   		The full kernel Eq.~\eqref{GammaFull} includes an additional Gaussian
   		suppression at separations exceeding the wavepacket coherence length
   		$\ell_c = (1+4\sigma_0^2\sigma_p^2)^{1/2}/\sigma_p$, so the
   		quoted regimes represent upper estimates. In the broad momentum factorized
   		limit $\sigma_p\to\infty$ the rate scales
   		as $\Gamma_N\propto N^2(m_e/m_f)^2 N_g$, where $N_g$ is the total graviton
   		number and $m_f\simeq m_p$ for composite matter. This leads to negligible
   		decoherence for microscopic systems and rapid classicalization for macroscopic
   		objects.%
   	}
   	\label{table}
   \end{table}
   
   \begin{figure}[tb]
   	\centering
   	\includegraphics[scale=1]{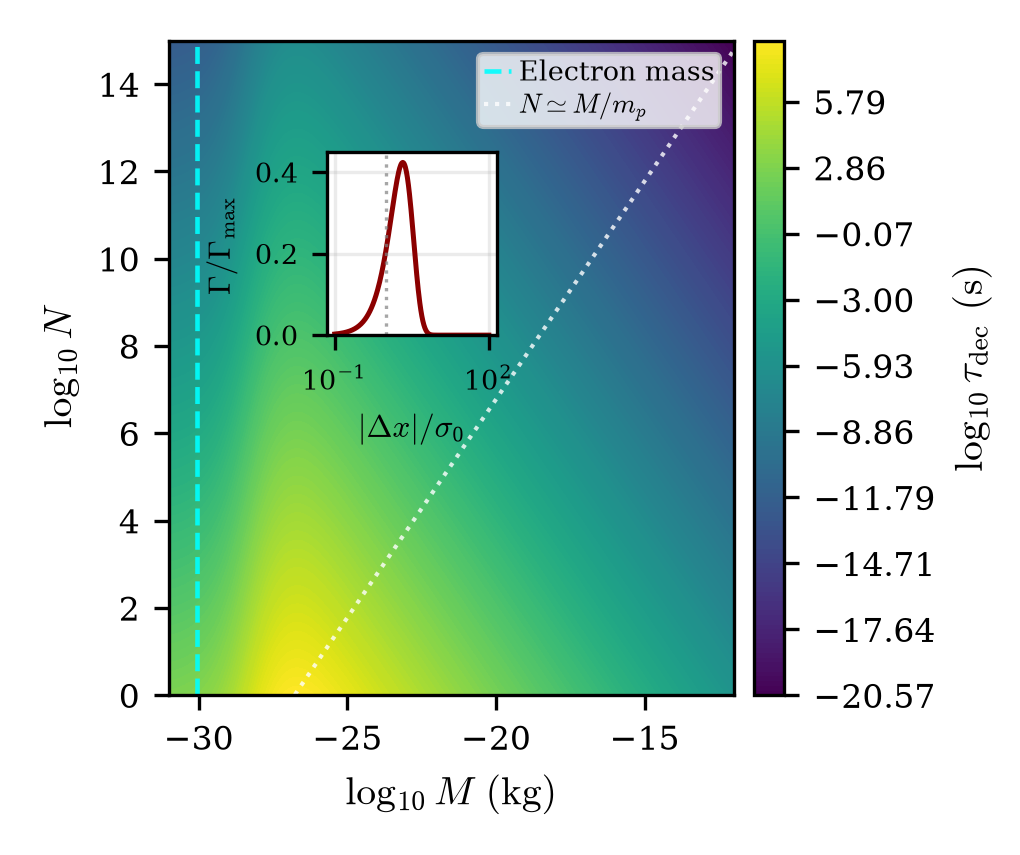}
   	\caption{%
   		Contour plot of the gravitationally induced decoherence time
   		$\tau_{\mathrm{dec}}=\Gamma^{-1}$ as a function of the total mass $M$ and the
   		constituent count $N$. The main panel incorporates the full decoherence kernel
   		Eq.~\eqref{GammaFull} evaluated at a representative spatial separation
   		$|\Delta\mathbf{x}|=\sigma_0$, and includes the Gaussian momentum width
   		suppression. The overall rate scales coherently as $\Gamma\propto N_{\mathrm{eff}}^2$
   		in the long wavelength regime, balanced by the constituent mass suppression
   		factor $(m_e/m_N)^2$ for composite nucleonic matter. The dashed cyan vertical
   		line marks the single electron mass ($M\simeq m_e$), highlighting the regime
   		where microscopic quantum coherence is preserved over cosmological timescales.
   		The dotted white line denotes physically realized compositions satisfying
   		$N\simeq M/m_p$. \textit{Inset:} The ratio $\Gamma/\Gamma_{\max}$ as a
   		function of the dimensionless separation $|\Delta x|/\sigma_0$, illustrating
   		the exact behavior of the full kernel: the rate vanishes at zero separation
   		(no which path information), peaks at intermediate separations, and is
   		exponentially suppressed at large separations due to the finite wavepacket
   		width.%
   	}
   	\label{fig1}
   \end{figure}

   \subsubsection{Hierarchy of physical regimes}
   
   The quadratic scaling in Eq.~\eqref{GammaN} generates a clear hierarchy across physical
   systems. In the idealized factorized limit, and for large spatial separation
   $|\Delta\mathbf{x}|\gg p_c^{-1}$ where $\operatorname{sinc}(p|\Delta\mathbf{x}|)\to0$,
   the decoherence kernel simplifies to
   \begin{equation}
   	\Gamma(\Delta\mathbf{x})
   	\approx
   	\frac{\kappa^2}{16m_f^2}
   	\mathcal{G}(\Delta\mathbf{x})\,
   	V
   	\int_0^\infty\frac{p^2\,dp}{2\pi^2}\,\tilde{I}^{(g)}(p)
   	=
   	\frac{\kappa^2}{16m_f^2}
   	\mathcal{G}(\Delta\mathbf{x})\,
   	N_g~.
   \end{equation}
   This expression should be understood as the maximal factorized estimate; the full kernel
   Eq.~\eqref{GammaFull} replaces the bracket $1-\sin(p|\Delta\mathbf{x}|)/(p|\Delta\mathbf{x}|)$
   by the Gaussian kernel $\mathcal{J}_{\sigma_p}$, which suppresses the rate at large
   separations for finite momentum width. Assuming an exponential spectrum
   \begin{equation}
   	\tilde{I}^{(g)}(p)
   	=
   	I_0\,e^{-p/p_c}~,
   \end{equation}
   the momentum integral yields
   $\int_0^\infty \frac{p^2\,dp}{2\pi^2}\,I_0\,e^{-p/p_c}=I_0p_c^3/\pi^2$, which leads to
   \begin{equation}
   	N_g
   	=
   	\frac{V I_0 p_c^3}{\pi^2}
   	\sim
   	V I_0 p_c^3~,
   \end{equation}
   showing explicitly that the rate is controlled by the total graviton number in the
   factorized limit.
   
   For $\sigma_0\sim10^{-9}\,\mathrm{m}$, one has
   $p_c\sim\hbar/\sigma_0\sim10^{-25}\,\mathrm{kg\,m/s}$, leading to a strong suppression
   of high momentum contributions at microscopic scales. The spatial kernel
   $\mathcal{G}(\Delta\mathbf{x})$ remains of order unity for separations
   $\Delta x\sim\sigma_0$.
   
   Using $\sigma_0\sim10^{-9}\,\mathrm{m}$ and $\Delta x\sim10^{-9}\,\mathrm{m}$, and
   working in the factorized limit, we obtain the maximal upper bounds
   \begin{align}
   	\Gamma_e^{\max}       &\sim 10^{-2}\,\mathrm{Hz} \notag\\
   	\Gamma_{\mathrm{mol}}^{\max}   &\sim 10^{-5}\,\mathrm{Hz}
   	\quad (N\sim10^2)~, \\
   	\Gamma_{\mathrm{virus}}^{\max} &\sim 10^{2}\text{--}10^{3}\,\mathrm{Hz}
   	\quad (N\sim10^6)~, \notag
   \end{align}
   for single electrons ($m_f=m_e$), small molecules, and larger composite systems such as
   viruses, respectively. These numbers represent upper estimates; the full kernel
   Eq.~\eqref{GammaFull} can only reduce them for finite momentum width $\sigma_p$,
   especially at larger separations. For composite neutral matter the emission is dominated
   by nucleons ($m_f\simeq m_p$), introducing a mass suppression factor
   $(m_e/m_p)^2\sim10^{-7}$ in Eq.~\eqref{GammaN}, so that the collective $N^2$ factor
   is required to overcome the nucleon mass suppression in the molecular and viral regimes.
   
   The resulting gravitationally induced decoherence time $\tau_{\mathrm{dec}}=\Gamma^{-1}$
   as a function of mass and constituent number is illustrated in Fig.~\ref{fig1}. The plot
   incorporates the full kernel evaluated at a representative separation
   $|\Delta\mathbf{x}|=\sigma_0$, showing how the Gaussian suppression modulates the
   maximal factorized scaling. The decoherence rate scales quadratically with the
   constituent number, $\Gamma\propto N^2$, reflecting coherent graviton emission, while
   being modulated by the constituent mass suppression $(m_e/m_N)^2$ in composite systems.
   This collective mechanism provides a microscopic route to classicality: while
   single particle effects remain negligible, coherent emission drives rapid suppression of
   quantum coherence in macroscopic matter.
   
   In Table~\ref{table}, we have summarized the resulting hierarchy using the maximal
   factorized estimates. Microscopic systems such as atoms and electrons correspond to
   $N\sim1$; no collective enhancement occurs and graviton induced decoherence is negligible,
   consistent with atomic interferometry experiments.
   For large molecules ($M\sim10^3$--$10^5~\mathrm{amu}$), one has $N\sim10^3$--$10^5$
   giving $N^2\lesssim10^{10}$; despite this enhancement the overall rate remains small due
   to the suppressed single particle coupling, consistent with observed matter wave
   interference.
   In the mesoscopic regime (nanoparticles with
   $M\sim10^{-18}$--$10^{-15}\,\mathrm{kg}$), one finds $N\sim10^9$--$10^{12}$ and
   $N^2\sim10^{18}$--$10^{24}$; amplification becomes significant and interference
   visibility is strongly reduced unless the system is extremely well isolated.
   For macroscopic systems ($N\gtrsim10^{15}$), the amplification becomes enormous with
   $N^2\gtrsim10^{30}$, so that even with extremely small $\Gamma_1$
   \begin{equation}
   	\tau_{\mathrm{dec}}^{(N)}\rightarrow0~,
   \end{equation}
   implying rapid suppression of quantum coherence and the emergence of classical behavior.

 \section{Conclusion}
\label{sec:conclusion}

In this work, we have developed a fully microscopic and field theoretic description of gravity induced decoherence based on graviton emission within the QBE framework. This approach provides a first principles realization of the idea that gravitational interactions can destabilize quantum superpositions without introducing phenomenological modifications to quantum mechanics.

Starting from the fermion graviton interaction, we derived a Markovian master equation for the reduced density matrix and showed that the off diagonal elements exhibit pure dephasing. In this picture, graviton emission acts as an intrinsic environmental monitoring channel that suppresses phase coherence while leaving populations unaffected. The resulting decoherence rate depends on the graviton spectral density, the normalization volume, and crucially, on the spatial structure and momentum width of the wavepacket. The normalization volume plays an important role, as it converts the spectral density into an extensive quantity proportional to the total number of accessible graviton modes, thereby linking decoherence to the effective graviton number.

A central result of this work is the derivation and analysis of the full decoherence kernel for a fermionic superposition with a finite momentum spread. Unlike the idealized factorized treatment—which yields a rate that monotonically increases with separation and saturates to a maximum value—the full kernel reveals a distinct phenomenology. The rate vanishes exactly at zero separation \(\Gamma(0)=0\) (since coincident branches carry no which path information), peaks at intermediate separations set by the wavepacket coherence length, and is exponentially suppressed as \(|\Delta\mathbf x|\to\infty\). This Gaussian suppression arises because, once the superposed branches are separated beyond the effective wavepacket width, the emitted radiation can no longer resolve the phase difference. This large distance cutoff sharply distinguishes graviton bremsstrahlung from static gravitational self energy mechanisms (such as the DP model) and CSL models, which predict either a slow power law decay or a saturation to a finite floor. 

Furthermore, we identified a coherent many body enhancement mechanism. In the long wavelength regime, where emitted gravitons cannot resolve the internal structure of the source, emission amplitudes from individual constituents add coherently. As a result, the decoherence rate grows quadratically with the number of constituents, and equivalently with the square of the total mass, modulated by the constituent mass suppression factor for composite nucleonic matter. This provides a direct and quantitative connection between microscopic dynamics and macroscopic behavior.

Our analysis reveals a clear hierarchy of regimes across the maximal factorized estimates. For microscopic systems such as electrons and atoms, gravitational decoherence is extremely weak. The small coupling strength, combined with the absence of collective enhancement, ensures that coherence times far exceed experimental timescales, fully consistent with the robustness of quantum interference observed in atomic and particle systems. For mesoscopic systems, including large molecules and nanoparticles, the quadratic amplification begins to compete with the smallness of the gravitational coupling, making decoherence increasingly sensitive to both system size and environmental graviton fluctuations. Finally, for macroscopic systems, the collective enhancement dominates. Even though the single particle coupling remains extremely weak and the finite width kernel suppresses long distance coherence loss, the large number of constituents leads to an effectively instantaneous suppression of quantum coherence. In this regime, superpositions become dynamically unstable, and classical behavior emerges naturally from the underlying quantum dynamics.

Ultimately, our results establish graviton bremsstrahlung as a continuous environmental monitoring channel. The interplay between the coherent \(N^2\) amplification and the wavepacket dependent spatial cutoff provides a theoretically robust and experimentally distinguishable signature of quantum gravity induced decoherence.

\section*{Acknowledgments}

MZ would like to thank the Department of Physics and Astronomy “G. Galilei” at the University of Padova for its warm hospitality during the completion of this work. He also thanks M. Abdi for very useful discussions and comments.

\section*{Declarations}

\subsection*{Funding and/or Conflicts of interests/Competing interests}
The authors declare that they have no competing interests.

\section*{Data Availability Statement}
No data are associated with this manuscript.

\appendix

\section{Calculation of the dissipator term due to graviton emission}

In this section, we provide the detailed calculation of the collision or damping term due to the bremsstrahlung emission described by Hamiltonian \eqref{Hint-2}. The dissipator term in the QBE \eqref{QBE1} is defined as
\begin{equation}
	D_{ij}[\rho^{(f)}]
	=
	-
	\int_0^t dx_1^0
	\left\langle
	\left[
	H_{\mathrm{int}}(x_1^0),
	\left[
	H_{\mathrm{int}}^\dagger(0),
	\hat{\mathcal D}_{ij}(\k)
	\right]
	\right]
	\right\rangle .
	\label{D1}
\end{equation}
Inserting the Hamiltonian \eqref{Hint-2} into the dissipator \eqref{QBE1-1} and using the expectation values defined as in
\cite{Kosowsky:1994cy,Bavarsad:2009hm,Bartolo:2018igk,Bartolo:2019eac,Hoseinpour:2020hic}, we get
\begin{align}
	D_{ij}[\rho^{(f)}]
	={}&
	-\left(\frac{\kappa}{2}\right)^2 V^2
	\sum_{s_1,r_1,r'_1}\sum_{s_2,r_2,r'_2}
	\int dx_1^0\,d\tau_1\,d\tau_2\,
	d\p_1\,d\q_1\,d\q'_1\,d\p_2\,d\q_2\,d\q'_2\,
	\delta(x_1^0-\bar x_1^0(\tau_1))\,
	\delta(x_2^0-\bar x_2^0(\tau_2))\big|_{x_2^0=0}
	\nonumber\\
	&\times
	h^{*s_1\mu_1\nu_1}(\p_1)\,
	q'_{1\mu_1}\,
	\bar u_{r'_1}(\q'_1)\gamma_{\nu_1}u_{r_1}(\q_1)\;
	h^{s_2\mu_2\nu_2}(\p_2)\,
	q'_{2\mu_2}\,
	\bar u_{r_2}(\q_2)\gamma_{\nu_2}u_{r'_2}(\q'_2)
	\nonumber\\
	&\times
	e^{-\sigma_0^2\left(|\q'_1-\q_1+\p_1|^2+|\q'_2-\q_2+\p_2|^2\right)}
	e^{-i(\q'_1-\q_1+\p_1)\cdot\bar{\x}_1(\tau_1)}
	e^{+i(\q'_2-\q_2+\p_2)\cdot\bar{\x}_2(\tau_2)}
	\nonumber\\
	&\times
	e^{i(q_1'^0-q_1^0+p_1^0)x_1^0}\,
	\rho^{(g)}_{s_2s_1}(\p_2)\,\delta^3(\p_1-\p_2)
	\nonumber\\
	&\times
	\Big[
	\delta_{r_1r_2}\delta_{r'_2 i}\,
	\delta^3(\q_2-\q_1)\delta^3(\k-\q'_2)\delta^3(\q'_1-\k)\,
	\rho^{(f)}_{jr'_1}(\k)
	\nonumber\\
	&\qquad
	-\delta_{r_1 i}\delta_{r_2 j}\,
	\delta^3(\k-\q_1)\delta^3(\q_2-\k)\delta^3(\q'_1-\q'_2)\,
	\rho^{(f)}_{r'_2r'_1}(\q'_2)
	\Big]~.
	\label{Dij-step1}
\end{align}

In the nonrelativistic limit, we approximate
\(\bar x_i^0(\tau_i)\simeq\tau_i\),
\(\bar{\x}_i(\tau_i)\simeq\bar{\x}_i\),
and \(q'^0\simeq q^0\simeq m_f\).
Using the temporal delta functions to perform the \(x_1^0\) and \(\tau_2\) integrations, the dissipator term reduces to
\begin{align}
	D_{ij}[\rho^{(f)}]
	={}&
	-\frac{1}{2}
	\left(\frac{\kappa}{2}\right)^2
	(2\pi)^3V^2
	\sum_{s_1,r_1,r'_1}
	\sum_{s_2,r_2,r'_2}
	\int d\tau_1\,d\p_1\,d\q_1\,d\q'_1\,
	d\p_2\,d\q_2\,d\q'_2\,
	e^{ip_1^0\tau_1}
	\nonumber\\
	&\times
	h^{s_1\mu_1\nu_1}(\p_1)
	\bar u_{r'_1}(\q'_1)
	(iq'_{1\mu_1})
	\gamma_{\nu_1}
	u_{r_1}(\q_1)\;
	h^{s_2\mu_2\nu_2}(\p_2)
	\bar u_{r'_2}(\q_2)
	(-iq'_{2\mu_2})
	\gamma_{\nu_2}
	u_{r_2}(\q'_2)
	\nonumber\\
	&\times
	\exp\!\left[-\sigma_0^2|\q'_1-\q_1+\p_1|^2\right]
	\exp\!\left[-i(\q'_1-\q_1+\p_1)\cdot\bar{\x}_1\right]
	\exp\!\left[-\sigma_0^2|\q'_2-\q_2+\p_2|^2\right]
	\exp\!\left[i(\q'_2-\q_2+\p_2)\cdot\bar{\x}_2\right]
	\nonumber\\
	&\times
	\rho^{(g)}_{s_2s_1}(\p_2)\,
	\delta^3(\p_1-\p_2)
	\Big[
	\delta_{r_1r_2}\delta_{r'_2i}
	\delta^3(\q_2-\q_1)
	\delta^3(\k-\q'_2)
	\delta^3(\q'_1-\k)
	\rho^{(f)}_{jr'_1}(\k)
	\nonumber\\
	&\qquad
	-
	\delta_{r_1i}\delta_{r_2j}
	\delta^3(\k-\q_1)
	\delta^3(\q_2-\k)
	\delta^3(\q'_1-\q'_2)
	\rho^{(f)}_{r'_2r'_1}(\q'_2)
	\Big]~.
	\label{Dij-step2}
\end{align}
After performing the momentum integrations using the delta functions, the collision term reduces to
\begin{align}
	D_{ij}[\rho^{(f)}]
	={}&
	-\frac{1}{2}
	\left(\frac{\kappa}{2}\right)^2
	(2\pi)^3V^2
	\sum_{s,s'}\sum_{r,r',r''}
	\int d\tau\,d\p\,d\q\,
	h^{s\mu\nu}(\p)
	h^{s'\mu'\nu'}(\p)
	\rho^{(g)}_{s's}(\p)
	\nonumber\\
	&\times
	\Bigg[
	\bar u_{r'}(\k)(ik_\mu)\gamma_\nu
	u_r(\q)\,
	\bar u_{r}(\q)(-ik_{\mu'})\gamma_{\nu'}
	u_i(\k)\,
	e^{i(k^0-q^0+p^0)\tau}
	\nonumber\\
	&\qquad\times
	e^{-\sigma_0^2|\k-\q+\p|^2}
	e^{-i(\k-\q+\p)\cdot(\bar{\x}_1-\bar{\x}_2)}
	\rho^{(f)}_{jr'}(\k)
	\nonumber\\
	&\quad-
	\bar u_{r'}(\q)(iq_\mu)\gamma_\nu
	u_i(\k)\,
	\bar u_{r''}(\k)(-iq_{\mu'})\gamma_{\nu'}
	u_j(\q)\,
	e^{i(q^0-k^0+p^0)\tau}
	\nonumber\\
	&\qquad\times
	e^{-\sigma_0^2|\q-\k+\p|^2}
	e^{-i(\q-\k+\p)\cdot(\bar{\x}_1-\bar{\x}_2)}
	\rho^{(f)}_{r''r'}(\q)
	\Bigg]~.
	\label{Dij-step3}
\end{align}

Using the Gordon identity
\begin{equation}
	\bar u(p')\gamma^\mu u(p)
	=
	\bar u(p')
	\left[
	\frac{(p'+p)^\mu}{2m_f}
	+
	\frac{i\sigma^{\mu\nu}(p'-p)_\nu}{2m_f}
	\right]
	u(p)~,
\end{equation}
to leading order, one finds
\begin{equation}
	\bar u_{r'}(\k)\gamma^\mu u_r(\q)
	=
	\delta^{\mu0}\chi_{r'}^\dagger\chi_r
	+
	\frac{1}{2m_f}
	\chi_{r'}^\dagger
	\left[
	(k+q)^j\delta^\mu_j
	+
	i\sigma^{\mu\nu}(k-q)_\nu
	\right]
	\chi_r ~.
\end{equation}
We define the effective nonrelativistic vertex operator
\begin{equation}
	\Gamma^{\mu\nu}(\k,\q)
	=
	i k^\mu
	\left[
	\delta^{\nu0}
	+
	\frac{1}{2m_f}
	\left(
	(k+q)^j\delta_j^\nu
	+
	i\sigma^{\nu\alpha}(k-q)_\alpha
	\right)
	\right]~.
	\label{Gammavertex}
\end{equation}
Then, directly in terms of \(\Gamma^{\mu\nu}\), the dissipator takes the form
\begin{align}
	D_{ij}[\rho^{(f)}]
	={}&
	-\frac{1}{2}
	\left(\frac{\kappa}{2}\right)^2
	(2\pi)^3V^2
	\sum_{s,s'}\sum_{r,r',r''}
	\int d\tau\,d\p\,d\q\;
	h^{s\mu\nu}(\p)
	h^{s'\mu'\nu'}(\p)
	\rho^{(g)}_{s's}(\p)
	\nonumber\\
	&\times
	\Bigg[
	\left(
	\chi_{r'}^\dagger
	\Gamma_{\mu\nu}(\k,\q)
	\chi_r
	\right)
	\left(
	\chi_{r}^\dagger
	\Gamma_{\mu'\nu'}(\q,\k)
	\chi_i
	\right)
	e^{i(k^0-q^0+|\p|)\tau}
	e^{-\sigma_0^2|\k-\q+\p|^2}
	e^{-i(\k-\q+\p)\cdot\Delta\x}
	\rho^{(f)}_{jr'}(\k)
	\nonumber\\
	&\qquad-
	\left(
	\chi_{r'}^\dagger
	\Gamma_{\mu\nu}(\q,\k)
	\chi_i
	\right)
	\left(
	\chi_{j}^\dagger
	\Gamma_{\mu'\nu'}(\k,\q)
	\chi_{r''}
	\right)
	e^{i(q^0-k^0+|\p|)\tau}
	e^{-\sigma_0^2|\q-\k+\p|^2}
	e^{-i(\q-\k+\p)\cdot\Delta\x}
	\rho^{(f)}_{r''r'}(\q)
	\Bigg]~,
	\label{Dij-Gamma}
\end{align}
where \(\Delta\x=\bar{\x}_1-\bar{\x}_2\).

 We now take the integration over the internal fermionic momentum \(\q\). The local contribution is obtained by
 \begin{equation}
 	\mathcal G(\Delta\x)
 	=
 	\int\frac{d^3\q}{(2\pi)^3}
 	\exp\!\left[-2\sigma_0^2|\k-\q|^2\right]
 	\exp\!\left[-i(\k-\q)\cdot\Delta\x\right]
 	=
 	\frac{1}{(8\pi\sigma_0^2)^{3/2}}
 	\exp\!\left(
 	-\frac{|\Delta\x|^2}{8\sigma_0^2}
 	\right)~.
 	\label{mathG}
 \end{equation}
 Physically, \(\sigma_0\) sets the coherence scale of the system,
 \(\ell_{\mathrm{coh}}\sim\sigma_0\), and ensures that the dissipative dynamics remains ultraviolet finite. Importantly, the integration over \(\q\) generates the local kernel \(\mathcal G(\Delta\x)\), while the graviton momentum \(\p\) remains in the nonlocal phase factor \(e^{-i\p\cdot\Delta\x}\), which is responsible for spatial decoherence after angular averaging.
 
 To retain the finite momentum spread of the fermionic wavepacket, we do not simply set \(\rho^{(f)}_{r''r'}(\q)=\rho^{(f)}_{r''r'}(\k)\). Instead, in the term where \(\rho^{(f)}\) is evaluated at \(\q\), we use the Gaussian ansatz
 \begin{equation}
 	\rho^{(f)}_{r''r'}(\q)
 	=
 	\rho^{(f)}_{r''r'}(\k)
 	\exp\!\left[
 	-\frac{|\q-\k|^2}{2\sigma_p^2}
 	\right]~,
 	\label{Gaussianrho-app}
 \end{equation}
 where \(\sigma_p\) is the momentum space width. With this profile, the corresponding \(\q\) integral in the dissipator is
 \begin{align}
 	\mathcal J_{\sigma_p}(\p,\Delta\x)
 	\equiv{}&
 	\int\frac{d^3\q}{(2\pi)^3}
 	\exp\!\left[-2\sigma_0^2|\k-\q+\p|^2\right]
 	\exp\!\left[-i(\k-\q+\p)\cdot\Delta\x\right]
 	\exp\!\left[-\frac{|\q-\k|^2}{2\sigma_p^2}\right]
 	\nonumber\\
 	={}&
 	\frac{1}{8\pi^{3/2}}
 	\left(
 	\frac{2\sigma_p^2}{1+4\sigma_0^2\sigma_p^2}
 	\right)^{3/2}
 	\exp\!\left[
 	-\frac{2\sigma_0^2p^2}{1+4\sigma_0^2\sigma_p^2}
 	-\frac{\sigma_p^2|\Delta\x|^2}{2(1+4\sigma_0^2\sigma_p^2)}
 	-\frac{i\,\p\cdot\Delta\x}{1+4\sigma_0^2\sigma_p^2}
 	\right]~.
 	\label{calJ}
 \end{align}
 The factor \(e^{-\sigma_p^2|\Delta\x|^2/2(1+4\sigma_0^2\sigma_p^2)}\) is the overlap of the initial wavepacket with its displaced copy; it already belongs to the initial coherence \(\rho^{(f)}_{12}\) and must not be counted again as a suppression of the decoherence rate itself. If the \(\q\) dependence of the phase and of the vertex is neglected, the integral reduces to the simple overlap factor
 \begin{equation}
 	\mathcal J_{\sigma_p}(\p,\Delta\x)
 	\simeq
 	\mathcal G(\Delta\x)\,
 	e^{-i\p\cdot\Delta\x}~,
 	\label{Japprox}
 \end{equation}
 which holds for \(p\sigma_0\ll1\) and \(\sigma_p\ll1/\sigma_0\), and from which the standard bracket \(1-\sin(p|\Delta\x|)/(p|\Delta\x|)\) is recovered after angular averaging. Only the cross term carries \(\mathcal G(\Delta\x)\) and the phase \(e^{-i\p\cdot\Delta\x}\), since the self emission terms have both vertices on the same worldline (\(\Delta\x=0\)).
 
 Under the soft graviton limit
 \(\rho^{(f)}_{r'i}(\k-\p)\approx\rho^{(f)}_{r'i}(\k)\)
 and
 \(\Gamma_{\mu'\nu'}(\k-\p)\approx\Gamma_{\mu'\nu'}(\k)\)~,
 the dissipator reduces to
 \begin{align}
 	D_{ij}[\rho^{(f)}]
 	={}&
 	-\frac{1}{2}
 	\left(\frac{\kappa}{2}\right)^2 V
 	\sum_{s,s'}\sum_{r,r'}
 	\int d\tau\,d^3\p\,
 	e^{-i|\p|\tau}
 	h^{s\mu\nu}(\p)
 	h^{s'\mu'\nu'}(\p)
 	\rho^{(g)}_{ss'}(\p)
 	\mathcal G(\Delta\x)
 	\nonumber\\
 	&\times
 	\Bigg[
 	\left(
 	\chi_{r'}^\dagger
 	\Gamma_{\mu\nu}(\k)
 	\Gamma_{\mu'\nu'}(\k)
 	\chi_i
 	\right)
 	\rho^{(f)}_{jr'}(\k)
 	-
 	\left(
 	\chi_j^\dagger
 	\Gamma_{\mu'\nu'}(\k)
 	\Gamma_{\mu\nu}(\k)
 	\chi_{r'}
 	\right)
 	\rho^{(f)}_{r'i}(\k)\,
 	e^{-i\p\cdot\Delta\x}
 	\Bigg]~,
 	\label{Dij-final}
 \end{align}
 where the factor \((2\pi)^3\delta^{(3)}(0)=V\) from the mode normalization reduces the overall volume scaling to linear order in \(V\).

 We now turn to integrate over the graviton momentum. For an unpolarized graviton state \eqref{Ig},
 \begin{equation}
 	\rho^{(g)}_{ss'}=\frac{1}{2}I^{(g)}(p)\delta_{ss'}~,
 \end{equation}
 and summing over physical TT polarizations yields
 \begin{equation}
 	\sum_s h^s_{ij}(\hat{\p})h^s_{kl}(\hat{\p})
 	=
 	\Pi^{TT}_{ij,kl}(\hat{\p})~.
 \end{equation}
 Angular integration gives the isotropic tensor identity
 \begin{equation}
 	\int d\Omega_{\hat{\p}}\,\Pi^{TT}_{ij,kl}
 	=
 	\frac{8\pi}{5}
 	\left(
 	\delta_{ik}\delta_{jl}
 	+
 	\delta_{il}\delta_{jk}
 	-
 	\frac{2}{3}\delta_{ij}\delta_{kl}
 	\right)~.
 \end{equation}
 The finite interaction time produces the exact filter function
 \begin{equation}
 	\int_0^{\tau_0}d\tau\,e^{-ip\tau}
 	=
 	e^{-ip\tau_0/2}
 	\frac{2\sin(p\tau_0/2)}{p}~.
 \end{equation}
 The graviton propagation direction is
 \begin{equation}
 	\hat{\p}
 	=
 	(\sin\theta\cos\phi,\;\sin\theta\sin\phi,\;\cos\theta)~.
 \end{equation}
 The polarization vectors are
 \begin{align}
 	e^{(1)}(\hat{\p})
 	&=
 	(\cos\theta\cos\phi,\;\cos\theta\sin\phi,\;-\sin\theta)~,
 	\\
 	e^{(2)}(\hat{\p})
 	&=
 	(-\sin\phi,\;\cos\phi,\;0)~.
 \end{align}
 The polarization tensors are therefore
 \begin{equation}
 	h^{(s)}_{ij}(\hat{\p})=e^{(s)}_i e^{(s)}_j,
 	\qquad s=1,2.
 \end{equation}
 We define the exact angular tensor
 \begin{equation}
 	\mathcal H_{ij,kl}(\theta,\phi)
 	=
 	\sum_{s=1}^{2}
 	h^{(s)}_{ij}(\theta,\phi)\,
 	h^{(s)}_{kl}(\theta,\phi)
 	=
 	\Pi^{TT}_{ij,kl}(\hat{\p})~,
 \end{equation}
 which is the explicit angular form of the TT polarization sum introduced above. Crucially, when evaluating the spin dependent interaction terms involving the Levi-Civita contraction
 \(\sum_{s=1}^2 h^{(s)}_{ij}(\theta,\phi)\epsilon_{ijm}\),
 we find that these terms vanish identically owing to the symmetric nature of the graviton polarization tensors \(h^{(s)}_{ij}=h^{(s)}_{ji}\) contrasted with the total antisymmetry of \(\epsilon_{ijm}\). Consequently, spin orbit cross couplings drop out entirely, leaving the non spin mass density sector as the sole surviving contribution driving spatial decoherence.
 
 Using completeness of Pauli spinors
 \begin{equation}
 	\sum_{r'=1}^2\chi_{r'}\chi_{r'}^\dagger=\mathbb I,
 \end{equation}
 we obtain full closure in spin space
 \begin{align}
 	\sum_{r'}(\chi_{r'}^\dagger A\chi_1)\rho_{1r'}
 	&=
 	(A\rho)_{11}~,
 	\\
 	\sum_{r'}(\chi_1^\dagger A\chi_{r'})\rho_{r'r}
 	&=
 	(\rho A^\dagger)_{11}~.
 \end{align}
 The non spin spatial interaction kernel simplifies after polarization sum to
 \begin{equation}
 	\mathcal W(\p,\theta,\phi)
 	=
 	\frac{\kappa^2V}{8m_f^2(2\pi)^3}
 	\mathcal F(\theta,\phi)
 	|\p|^2
 	\int d\tau\,e^{-i|\p|\tau}
 	I^{(g)}(|\p|)~,
 \end{equation}
 where the polarization factor is
 \begin{equation}
 	\mathcal F(\theta,\phi)=\sin^2\theta+\cos^2\theta\sin^2\phi~.
 \end{equation}
 Its isotropic angular average yields
 \(\int d\Omega_{\p}\,\mathcal F(\theta,\phi)=4\pi\).
 For the interference cross term, we use the standard spherical wave expansion
 \begin{equation}
 	\int d\Omega_{\p}\,\mathcal F(\theta,\phi)e^{-i\p\cdot\Delta\x}
 	=
 	4\pi\,\frac{\sin(p|\Delta\x|)}{p|\Delta\x|}~.
 \end{equation}
 Integrating $\mathcal W(\p,\theta,\phi)$ over the graviton momentum with these two angular averages produces, respectively, the local population kernel and the nonlocal coherence kernel entering $\Gamma(\Delta\x)$ below.
 
 Factoring out the shared spatial overlap kernel \(\mathcal G(\Delta\x)\) to maintain unitarity and trace preservation
 (\(\operatorname{Tr}[\dot\rho]=0\)), the integrated master equation takes the form
 \begin{equation}
 	\dot\rho_{11}(t)=0~,
 	\qquad
 	\dot\rho_{22}(t)=0~.
 \end{equation}
 Under the isotropic gravitational environment, the population sector remains stationary
 (\(\dot\rho_{11}=\dot\rho_{22}=0\)), demonstrating that spatial gravitational fluctuations cause pure dephasing without inducing state transitions or population decay.
 
 We now turn to the evolution of the off diagonal coherence \(\rho_{12}(\k,t)\). Unlike the diagonal elements, \(\rho_{12}\) encodes quantum phase information and is directly sensitive to environmental averaging. Performing the exact same angular integration for the off diagonal sector yields
 \begin{equation}
 	\dot\rho_{12}(t)
 	=
 	-\Gamma(\Delta\x)\rho_{12}(t)~,
 \end{equation}
 where the spatial decoherence kernel \(\Gamma(\Delta\x)\) is defined as
 \begin{equation}
 	\begin{aligned}
 		\Gamma(\Delta\x)
 		={}&
 		\frac{\kappa^2V}{16m_f^2}
 		\int_0^\infty\frac{p^2dp}{2\pi^2}\,
 		\tilde I^{(g)}(p)\,
 		\Bigg[
 		\mathcal G(\Delta\x)
 		-
 		\frac{1}{4\pi}
 		\int d\Omega_{\hat{\p}}\,
 		\mathcal F(\hat{\p})\,
 		\mathcal J_{\sigma_p}(\p,\Delta\x)
 		\Bigg]~,
 	\end{aligned}
 	\label{GammaFull}
 \end{equation}
 with \(\mathcal G(\Delta\x)\) given by Eq.~\eqref{mathG}, \(\mathcal J_{\sigma_p}\) given by Eq.~\eqref{calJ}, and
 \begin{equation}
 	\tilde I^{(g)}(p)
 	=
 	\int d\tau\,e^{-ip\tau}I^{(g)}(p)~.
 \end{equation}
 Equation~\eqref{GammaFull} is the full decoherence factor within the Gaussian momentum-profile ansatz~\eqref{Gaussianrho}. In the factorized approximation~\eqref{Japprox}, it reduces to the standard form
 \begin{equation}
 	\Gamma(\Delta\x)
 	=
 	\frac{\kappa^2V}{16m_f^2}
 	\mathcal G(\Delta\x)
 	\int_0^\infty\frac{p^2dp}{2\pi^2}
 	\tilde I^{(g)}(p)
 	\left[
 	1
 	-
 	\frac{\sin(p|\Delta\x|)}{p|\Delta\x|}
 	\right]~.
 	\label{Gamma}
 \end{equation}
 The evolution of the conjugate off-diagonal element satisfies
 \(\dot\rho_{21}(t)=(\dot\rho_{12}(t))^*=-\Gamma(\Delta\x)\rho_{21}(t)\).
 Thus, both off diagonal terms decay exponentially with rate \(\Gamma(\Delta\x)\).
 The isotropic gravitational background induces pure exponential dephasing on spatial superpositions while leaving energy populations invariant. The complete density matrix evolution under this pure dephasing channel drives the off diagonal coherences to zero as \(t\to\infty\), acting as a completely positive spatial dephasing channel.

\nocite{apsrev41Control} 
\bibliographystyle{unsrtnat}
\bibliography{Rev-refs}

@article{ManshouriZarei2025,
	author  = {Manshouri, Hassan and Zarei, Moslem},
	title   = {Dephasing through Bremsstrahlung emission: insights from quantum Boltzmann equation},
	journal = {The European Physical Journal Plus},
	volume  = {140},
	number  = {11},
	pages   = {1086},
	year    = {2025},
	doi     = {10.1140/epjp/s13360-025-07016-2},
	eprint  = {2510.18833},
	archivePrefix = {arXiv},
	primaryClass = {quant-ph}
}

@article{Sharifian2024,
	author  = {Sharifian, Mohammad and Zarei, Moslem and Abdi, Mehdi and Bartolo, Nicola and Matarrese, Sabino},
	title   = {Open quantum system approach to the gravitational decoherence of spin-1/2 particles},
	journal = {Physical Review D},
	volume  = {109},
	number  = {4},
	pages   = {043510},
	year    = {2024},
	doi     = {10.1103/PhysRevD.109.043510}
}

@article{DeLorenciFord:2015,
	author = {De Lorenci, V. A. and Ford, L. H.},
	title = {Decoherence induced by long wavelength gravitons},
	journal = {Phys. Rev. D},
	volume = {91},
	pages = {044038},
	year = {2015},
	doi = {10.1103/PhysRevD.91.044038},
	eprint = {1412.4685},
	archivePrefix = {arXiv},
	primaryClass = {gr-qc}
}

@article{Kanno:2021noise,
	author = {Kanno, Sugumi and Soda, Jiro and Tokuda, Junsei},
	title = {Noise and decoherence induced by gravitons},
	journal = {Phys. Rev. D},
	volume = {103},
	pages = {044017},
	year = {2021},
	doi = {10.1103/PhysRevD.103.044017}
}

@article{Toros:2024coherence,
	author = {Toro{\v{s}}, Marko and Mazumdar, Anupam and Bose, Sougato},
	title = {Loss of coherence and coherence protection from a graviton bath},
	journal = {Phys. Rev. D},
	volume = {109},
	number = {8},
	pages = {084050},
	year = {2024},
	doi = {10.1103/PhysRevD.109.084050}
}

@article{LintonTiwari:2025,
	author = {Linton, Flynn and Tiwari, Shubhanshu},
	title = {Gravitational waves decohere quantum superpositions},
	journal = {Class. Quantum Grav.},
	volume = {42},
	number = {20},
	pages = {205005},
	year = {2025},
	doi = {10.1088/1361-6382/ae0d27},
	eprint = {2501.18111},
	archivePrefix = {arXiv},
	primaryClass = {gr-qc}
}

@article{Matsui:2026,
	author = {Matsui, Hiroki},
	title = {Graviton-induced which-path decoherence in matter-wave interferometry},
	year = {2026},
	eprint = {2607.20867},
	archivePrefix = {arXiv},
	primaryClass = {gr-qc},
	doi = {10.48550/arXiv.2607.20867}
}

@article{Breuer:2001,
	author        = {Breuer, Heinz-Peter and Petruccione, Francesco},
	title         = {Destruction of quantum coherence through emission of bremsstrahlung},
	journal       = {Phys. Rev. A},
	volume        = {63},
	pages         = {032102},
	year          = {2001},
	month         = {Feb},
	doi           = {10.1103/PhysRevA.63.032102},
	eprint        = {},
	archivePrefix = {},
}

@article{Sharifian:2025olk,
	author = "Sharifian, Mohammad and Zarei, Moslem and Bartolo, Nicola and Matarrese, Sabino",
	title = "{Exploring the gravitational impulse via the quantum Boltzmann equation}",
	eprint = "2501.13678",
	archivePrefix = "arXiv",
	primaryClass = "gr-qc",
	doi = "10.1103/td49-w7pz",
	journal = "Phys. Rev. D",
	volume = "112",
	number = "6",
	pages = "064059",
	year = "2025"
}

@article{Zarei:2025nnm,
	author = "Zarei, Moslem and Abdi, Mehdi and Bartolo, Nicola and Matarrese, Sabino",
	title = "{Gravitationally mediated entanglement of fermionic qubits: from static to dynamical limits}",
	eprint = "2510.20587",
	archivePrefix = "arXiv",
	primaryClass = "quant-ph",
	doi = "10.1088/2058-9565/ae7468",
	journal = "Quantum Sci. Technol.",
	volume = "11",
	number = "3",
	pages = "035039",
	year = "2026"
}

@article{Sharifian:2023jem,
	author = "Sharifian, Mohammad and Zarei, Moslem and Abdi, Mehdi and Bartolo, Nicola and Matarrese, Sabino",
	title = "{Open quantum system approach to the gravitational decoherence of spin-1/2 particles}",
	eprint = "2309.07236",
	archivePrefix = "arXiv",
	primaryClass = "gr-qc",
	doi = "10.1103/PhysRevD.109.043510",
	journal = "Phys. Rev. D",
	volume = "109",
	number = "4",
	pages = "043510",
	year = "2024"
}

@article{Blencowe2013,
	author        = {Blencowe, M. P.},
	title         = {Effective Field Theory Approach to Gravitationally Induced Decoherence},
	journal       = {Physical Review Letters},
	volume        = {111},
	pages         = {021302},
	year          = {2013},
	doi           = {10.1103/PhysRevLett.111.021302},
	eprint        = {1211.4751},
	archivePrefix = {arXiv},
	primaryClass  = {quant-ph}
}

@article{AnastopoulosHu2013,
	author        = {Anastopoulos, C. and Hu, B. L.},
	title         = {A Master Equation for Gravitational Decoherence: Probing the Textures of Spacetime},
	journal       = {Classical and Quantum Gravity},
	volume        = {30},
	pages         = {165007},
	year          = {2013},
	doi           = {10.1088/0264-9381/30/16/165007},
	eprint        = {1305.5231},
	archivePrefix = {arXiv},
	primaryClass  = {gr-qc}
}

@misc{Riedel2013,
	author        = {Riedel, C. Jess},
	title         = {Evidence for Gravitons from Decoherence by Bremsstrahlung},
	year          = {2013},
	eprint        = {1310.6347},
	archivePrefix = {arXiv},
	primaryClass  = {quant-ph},
	doi           = {10.48550/arXiv.1310.6347}
}

@article{OnigaWang2017,
	author        = {Oniga, Teodora and Wang, Charles H.-T.},
	title         = {Quantum Coherence, Radiance, and Resistance of Gravitational Systems},
	journal       = {Physical Review D},
	volume        = {96},
	pages         = {084014},
	year          = {2017},
	doi           = {10.1103/PhysRevD.96.084014},
	eprint        = {1701.04122},
	archivePrefix = {arXiv},
	primaryClass  = {gr-qc}
}

@article{Diosi:1989hlx,
	author = "Di\'osi, L.",
	title = "{Models for universal reduction of macroscopic quantum fluctuations}",
	doi = "10.1103/PhysRevA.40.1165",
	journal = "Phys. Rev. A",
	volume = "40",
	number = "3",
	pages = "1165",
	year = "1989"
}

@article{Penrose:1996cv,
	author = "Penrose, Roger",
	title = "{On gravity's role in quantum state reduction}",
	doi = "10.1007/BF02105068",
	journal = "Gen. Rel. Grav.",
	volume = "28",
	pages = "581--600",
	year = "1996"
}

@article{Donadi:2020kzc,
	author = "Donadi, Sandro and Piscicchia, Kristian and Curceanu, Catalina and Di\'osi, Lajos and Laubenstein, Matthias and Bassi, Angelo",
	title = "{Underground test of gravity-related wave function collapse}",
	eprint = "2111.13490",
	archivePrefix = "arXiv",
	primaryClass = "quant-ph",
	doi = "10.1038/s41567-020-1008-4",
	journal = "Nature Phys.",
	volume = "17",
	number = "1",
	pages = "74--78",
	year = "2021"
}

@article{Piscicchia:2024wan,
	author = "Piscicchia, Kristian and Donadi, Sandro and Manti, Simone and Bassi, Angelo and Derakhshani, Maaneli and Di\'osi, Lajos and Curceanu, Catalina",
	title = "{X-Ray Emission from Atomic Systems Can Distinguish between Prevailing Dynamical Wave-Function Collapse Models}",
	doi = "10.1103/PhysRevLett.132.250203",
	journal = "Phys. Rev. Lett.",
	volume = "132",
	number = "25",
	pages = "250203",
	year = "2024"
}

@article{deSouza:2024ymq,
	author = "de Souza, Vitoria A. and Ribeiro, Caio C. Holanda and De Lorenci, Vitorio A.",
	title = "{Macroscopic quantum superpositions in superconducting circuits}",
	eprint = "2406.06492",
	archivePrefix = "arXiv",
	primaryClass = "quant-ph",
	month = "6",
	year = "2024"
}

@article{Tagg:2024fvq,
	author = "Tagg, James and Reid, William and Carlin, Daniel",
	title = {{Schr\"odinger's Cheshire Cat: A tabletop experiment to measure the Di\'osi-Penrose collapse time and demonstrate Objective Reduction (OR)}},
	eprint = "2402.02618",
	archivePrefix = "arXiv",
	primaryClass = "quant-ph",
	month = "2",
	year = "2024"
}

@article{DiBartolomeo:2024wav,
	author = "Di Bartolomeo, Giovanni and Carlesso, Matteo",
	title = "{Experimental bounds on linear-friction dissipative collapse models from levitated optomechanics}",
	eprint = "2401.04665",
	archivePrefix = "arXiv",
	primaryClass = "quant-ph",
	doi = "10.1088/1367-2630/ad3842",
	journal = "New J. Phys.",
	volume = "26",
	number = "4",
	pages = "043006",
	year = "2024"
}

@proceedings{Proceedings:2023mkp,
	author = "Abend, Sven and others",
	title = "{Terrestrial Very-Long-Baseline Atom Interferometry: Workshop Summary}",
	eprint = "2310.08183",
	archivePrefix = "arXiv",
	primaryClass = "hep-ex",
	doi = "10.1116/5.0185291",
	month = "10",
	year = "2023"
}

@article{ICECUBE:2023gdv,
	author = "Abbasi, R. and others",
	collaboration = "ICECUBE, IceCube",
	title = "{Search for decoherence from quantum gravity with atmospheric neutrinos}",
	eprint = "2308.00105",
	archivePrefix = "arXiv",
	primaryClass = "hep-ex",
	doi = "10.1038/s41567-024-02436-w",
	journal = "Nature Phys.",
	volume = "20",
	number = "6",
	pages = "913--920",
	year = "2024"
}

@article{Fadel:2023ici,
	author = "Fadel, Matteo",
	title = {{Probing gravity-related decoherence with a 16 $\mu$g Schr\"odinger cat state}},
	eprint = "2305.04780",
	archivePrefix = "arXiv",
	primaryClass = "quant-ph",
	month = "5",
	year = "2023"
}

@article{Napolitano:2023lar,
	author = "Napolitano, Fabrizio and others",
	title = "{Underground Tests of Quantum Mechanics by the VIP Collaboration at Gran Sasso}",
	doi = "10.3390/sym15020480",
	journal = "Symmetry",
	volume = "15",
	number = "2",
	pages = "480",
	year = "2023"
}

@article{Bassi:2022ibq,
	author = "Bassi, A. and others",
	title = "{A way forward for fundamental physics in space}",
	doi = "10.1038/s41526-022-00229-0",
	journal = "npj Microgravity",
	volume = "8",
	number = "1",
	pages = "49",
	year = "2022"
}

@article{Simonov:2022epo,
	author = "Simonov, Kyrylo",
	title = "{Observability of spontaneous collapse in flavor oscillations and its relation to the CP and CPT symmetries}",
	eprint = "2208.14383",
	archivePrefix = "arXiv",
	primaryClass = "hep-ph",
	doi = "10.1016/j.physleta.2022.128413",
	journal = "Phys. Lett. A",
	volume = "452",
	pages = "128413",
	year = "2022"
}

@article{Polkovnikov:2022hkg,
	author = "Polkovnikov, Mark and Gramolin, Alexander V. and Kaplan, David E. and Rajendran, Surjeet and Sushkov, Alexander O.",
	title = "{Experimental Limit on Nonlinear State-Dependent Terms in Quantum Theory}",
	eprint = "2204.11875",
	archivePrefix = "arXiv",
	primaryClass = "quant-ph",
	reportNumber = "FERMILAB-PUB-22-977-SQMS-V",
	doi = "10.1103/PhysRevLett.130.040202",
	journal = "Phys. Rev. Lett.",
	volume = "130",
	number = "4",
	pages = "040202",
	year = "2023"
}

@article{Curceanu:2022yeg,
	author = "Curceanu, Catalina and others",
	title = "{Underground tests of Quantum Mechanics at Gran Sasso}",
	doi = "10.22323/1.405.0005",
	journal = "PoS",
	volume = "DISCRETE2020-2021",
	pages = "005",
	year = "2022"
}

@article{Oppenheim:2022xjr,
	author = "Oppenheim, Jonathan and Sparaciari, Carlo and \v{S}oda, Barbara and Weller-Davies, Zachary",
	title = "{Gravitationally induced decoherence vs space-time diffusion: testing the quantum nature of gravity}",
	eprint = "2203.01982",
	archivePrefix = "arXiv",
	primaryClass = "quant-ph",
	doi = "10.1038/s41467-023-43348-2",
	journal = "Nature Commun.",
	volume = "14",
	number = "1",
	pages = "7910",
	year = "2023"
}

@article{Carlesso:2022pqr,
	author = "Carlesso, Matteo and Donadi, Sandro and Ferialdi, Luca and Paternostro, Mauro and Ulbricht, Hendrik and Bassi, Angelo",
	title = "{Present status and future challenges of non-interferometric tests of collapse models}",
	eprint = "2203.04231",
	archivePrefix = "arXiv",
	primaryClass = "quant-ph",
	doi = "10.1038/s41567-021-01489-5",
	journal = "Nature Phys.",
	volume = "18",
	number = "3",
	pages = "243--250",
	year = "2022"
}

@article{Gasbarri:2021sdm,
	author = "Gasbarri, Giulio and Belenchia, Alessio and Carlesso, Matteo and Donadi, Sandro and Bassi, Angelo and Kaltenbaek, Rainer and Paternostro, Mauro and Ulbricht, Hendrik",
	title = "{Testing the foundations of quantum physics in space Interferometric and non-interferometric tests with Large Particles}",
	eprint = "2106.05349",
	archivePrefix = "arXiv",
	primaryClass = "quant-ph",
	doi = "10.1038/s42005-021-00656-7",
	journal = "Commun. Phys.",
	volume = "4",
	pages = "155",
	year = "2021"
}

@article{Belenchia:2021rfb,
	author = "Belenchia, Alessio and others",
	title = "{Quantum physics in space}",
	eprint = "2108.01435",
	archivePrefix = "arXiv",
	primaryClass = "quant-ph",
	doi = "10.1016/j.physrep.2021.11.004",
	journal = "Phys. Rept.",
	volume = "951",
	pages = "1--70",
	year = "2022"
}

@article{Sigl:1993ctk,
	author = "Sigl, G. and Raffelt, G.",
	title = "{General kinetic description of relativistic mixed neutrinos}",
	reportNumber = "MPI-PH-92-112",
	doi = "10.1016/0550-3213(93)90175-O",
	journal = "Nucl. Phys. B",
	volume = "406",
	pages = "423--451",
	year = "1993"
}

@article{Kosowsky:1994cy,
	author = "Kosowsky, Arthur",
	title = "{Cosmic microwave background polarization}",
	eprint = "astro-ph/9501045",
	archivePrefix = "arXiv",
	reportNumber = "FERMILAB-PUB-94-206-A",
	doi = "10.1006/aphy.1996.0020",
	journal = "Annals Phys.",
	volume = "246",
	pages = "49--85",
	year = "1996"
}

@article{Bavarsad:2009hm,
	author = "Bavarsad, E. and Haghighat, M. and Rezaei, Z. and Mohammadi, R. and Motie, I. and Zarei, M.",
	title = "{Generation of circular polarization of the CMB}",
	eprint = "0912.2993",
	archivePrefix = "arXiv",
	primaryClass = "hep-th",
	doi = "10.1103/PhysRevD.81.084035",
	journal = "Phys. Rev. D",
	volume = "81",
	pages = "084035",
	year = "2010"
}

@article{Bartolo:2018igk,
	author = "Bartolo, Nicola and Hoseinpour, Ahmad and Orlando, Giorgio and Matarrese, Sabino and Zarei, Moslem",
	title = "{Photon-graviton scattering: A new way to detect anisotropic gravitational waves?}",
	eprint = "1804.06298",
	archivePrefix = "arXiv",
	primaryClass = "gr-qc",
	doi = "10.1103/PhysRevD.98.023518",
	journal = "Phys. Rev. D",
	volume = "98",
	number = "2",
	pages = "023518",
	year = "2018"
}

@article{Bartolo:2019eac,
	author = "Bartolo, Nicola and Hoseinpour, Ahmad and Matarrese, Sabino and Orlando, Giorgio and Zarei, Moslem",
	title = "{CMB Circular and B-mode Polarization from New Interactions}",
	eprint = "1903.04578",
	archivePrefix = "arXiv",
	primaryClass = "hep-ph",
	doi = "10.1103/PhysRevD.100.043516",
	journal = "Phys. Rev. D",
	volume = "100",
	number = "4",
	pages = "043516",
	year = "2019"
}

@article{Hoseinpour:2020hic,
	author = "Hoseinpour, Ahmad and Zarei, Moslem and Orlando, Giorgio and Bartolo, Nicola and Matarrese, Sabino",
	title = "{CMB $V$ modes from photon-photon forward scattering revisited}",
	eprint = "2006.14418",
	archivePrefix = "arXiv",
	primaryClass = "hep-ph",
	doi = "10.1103/PhysRevD.102.063501",
	journal = "Phys. Rev. D",
	volume = "102",
	number = "6",
	pages = "063501",
	year = "2020"
}

@article{Zarei:2021dpb,
	author = "Zarei, Moslem and Bartolo, Nicola and Bertacca, Daniele and Ricciardone, Angelo and Matarrese, Sabino",
	title = "{Non-Markovian open quantum system approach to the early Universe: Damping of gravitational waves by matter}",
	eprint = "2104.04836",
	archivePrefix = "arXiv",
	primaryClass = "astro-ph.CO",
	doi = "10.1103/PhysRevD.104.083508",
	journal = "Phys. Rev. D",
	volume = "104",
	number = "8",
	pages = "083508",
	year = "2021"
}

@article{Smirne:2014paa,
	author = "Smirne, Andrea and Bassi, Angelo",
	title = "{Dissipative Continuous Spontaneous Localization (CSL) model}",
	eprint = "1408.6446",
	archivePrefix = "arXiv",
	primaryClass = "quant-ph",
	doi = "10.1038/srep12518(2015)",
	journal = "Sci. Rep.",
	volume = "5",
	pages = "12518",
	year = "2015"
}

@article{Carlesso:2016khv,
	author = "Carlesso, M. and Bassi, A. and Falferi, P. and Vinante, A.",
	title = "{Experimental bounds on collapse models from gravitational wave detectors}",
	eprint = "1606.04581",
	archivePrefix = "arXiv",
	primaryClass = "quant-ph",
	doi = "10.1103/PhysRevD.94.124036",
	journal = "Phys. Rev. D",
	volume = "94",
	number = "12",
	pages = "124036",
	year = "2016"
}

@article{Bassi:2012bg,
	author = "Bassi, Angelo and Lochan, Kinjalk and Satin, Seema and Singh, Tejinder P. and Ulbricht, Hendrik",
	title = "{Models of Wave-function Collapse, Underlying Theories, and Experimental Tests}",
	eprint = "1204.4325",
	archivePrefix = "arXiv",
	primaryClass = "quant-ph",
	doi = "10.1103/RevModPhys.85.471",
	journal = "Rev. Mod. Phys.",
	volume = "85",
	pages = "471--527",
	year = "2013"
}

@article{Tilloy:2015zya,
	author = "Tilloy, Antoine and Di\'osi, Lajos",
	title = "{Sourcing semiclassical gravity from spontaneously localized quantum matter}",
	eprint = "1509.08705",
	archivePrefix = "arXiv",
	primaryClass = "quant-ph",
	doi = "10.1103/PhysRevD.93.024026",
	journal = "Phys. Rev. D",
	volume = "93",
	number = "2",
	pages = "024026",
	year = "2016"
}

@article{Kafri:2014zsa,
	author = "Kafri, D. and Taylor, J. M. and Milburn, G. J.",
	title = "{A classical channel model for gravitational decoherence}",
	eprint = "1401.0946",
	archivePrefix = "arXiv",
	primaryClass = "quant-ph",
	doi = "10.1088/1367-2630/16/6/065020",
	journal = "New J. Phys.",
	volume = "16",
	pages = "065020",
	year = "2014"
}




\end{document}